\documentclass[journal,twocolumn, twoside]{IEEEtran}
\usepackage[T1]{fontenc}

\usepackage{amsmath}
\usepackage{hyperref}
\usepackage{mathrsfs}
\usepackage{mathrsfs}
\usepackage{amsfonts}
\usepackage{graphicx,cite,epsfig,amssymb,amsmath}
\usepackage{color,xcolor}
\usepackage{bm}
\definecolor{red}{rgb}{1.00, 0.00, 0.00}  
\usepackage{pifont}
\usepackage{stmaryrd}
\usepackage{setspace}
\usepackage{subfigure}
\usepackage{cite}
\usepackage{array}
\usepackage{float}
\usepackage{multirow}
\usepackage{algorithm,algpseudocode}
\usepackage{epstopdf}
\usepackage{mathtools}
\usepackage{amssymb}
\usepackage{diagbox}
\usepackage{booktabs}
\usepackage{makecell}

\usepackage{xcolor}
\makeatletter

\newcommand{\Rmnum}[1]{\expandafter\@slowromancap\romannumeral #1@}
\makeatother
\usepackage{pifont} 
\usepackage{epstopdf}
\usepackage{mathtools}
\usepackage{diagbox}
\usepackage{booktabs}
\usepackage{multirow}
\usepackage{booktabs}
\usepackage{makecell}
\usepackage{float}
\usepackage{subfigure}
\usepackage{colortbl}

\newtheorem{Definition}{Definition}
\newtheorem{Theo}{Theorem}
\newtheorem{lemma}{Lemma}

\begin{document}
\normalsize
\title{Meta-Gating Framework for Fast and Continuous Resource Optimization in Dynamic Wireless Environments}
\author{\large Qiushuo Hou, Mengyuan Lee, Guanding Yu, and Yunlong Cai
\thanks{Q. Hou, M. Lee, G. Yu, and Y. Cai are with the College of Information Science and Electronic Engineering, Zhejiang University, Hangzhou 310027, China. e-mail: \{qshou, mengyuan\_lee, yuguanding, ylcai\}@zju.edu.cn.}}
\maketitle
\vspace{-1.5cm}
\begin{abstract}
\indent With the great success of deep learning (DL) in image classification, speech recognition, and other fields, more and more studies have applied various neural networks (NNs) to wireless resource allocation. Generally speaking, these artificial intelligent (AI) models are trained under some special learning hypotheses, especially that the statistics of the training data are static during the training stage. However, the distribution of channel state information (CSI) is constantly changing in the real-world wireless communication environment. Therefore, it is essential to study effective dynamic DL technologies to solve wireless resource allocation problems. In this paper, we propose a novel framework, named meta-gating, for solving resource allocation problems in an episodically dynamic wireless environment, where the CSI distribution changes over periods and remains constant within each period. The proposed framework, consisting of an inner network and an outer network, aims to adapt to the dynamic wireless environment by achieving three important goals, i.e., seamlessness, quickness and continuity. Specifically, for the former two goals, we propose a training method by combining a model-agnostic meta-learning (MAML) algorithm with an unsupervised learning mechanism. With this training method, the inner network is able to fast adapt to different channel distributions because of the good initialization. As for the goal of ‘continuity’, the outer network can learn to evaluate the importance of inner network’s parameters under different CSI distributions, and then decide which subset of the inner network should be activated through the gating operation. Additionally, we theoretically analyze the performance of the proposed meta-gating framework. Simulation results demonstrate that the proposed meta-gating framework can well achieve the three important goals compared with existing state-of-the-art algorithms.
\end{abstract}
\begin{IEEEkeywords}
Dynamic wireless environment, meta-learning, continual learning, resource allocation, neural network.
\end{IEEEkeywords}
\vspace{-0.3cm}
\section{Introduction}\label{introduction}

Resource allocation plays an essential role in wireless communications. However, most of them are formulated as NP-hard non-convex problems, which are computationally challenging to solve. With the great success of deep learning (DL) in image classification, speech recognition, and other fields, various neural networks (NNs) are recently applied to solve resource allocation problems in wireless networks\cite{DNN1, DNN2, CNN1, CNN2}. In \cite{DNN1} and \cite{DNN2}, the deep neural networks (DNNs) trained by the unsupervised learning method were employed to solve the power control problem for sum-rate maximization. The authors in \cite{CNN1} have designed a convolutional neural network (CNN) to optimize the transmit power in device-to-device (D2D) networks. Recently, graph neural networks (GNNs) have been widely applied to solve resource allocation problems because of their good representation ability for wireless networks\cite{GNN1, GNN2, GNN3, GNN4}. In \cite{GNN1}, a GNN trained by the unsupervised learning method was applied to address the link scheduling in D2D networks. The authors in \cite{GNN2} have developed a GNN to solve the beamformer design problem in the multi-antenna systems. In \cite{GNN3, GNN4}, GNNs were designed to optimally allocate resources across a set of transceiver pairs in a wireless network. However, all aforementioned NNs are trained under some special hypotheses, in particular that the statistics of the training data are static. Unfortunately, the real-world wireless environment is dynamic and constantly changing, such as the distribution of channel state information (CSI) may change over periods. It is known that the NN-based methods in existing works usually suffer from severe performance degradation when the environment changes, i.e., when the real-time data follows a different distribution from that used in the training phase\cite{dynamic}. Besides, if one chooses to retrain the entire NN once the environment changes, the re-training process would incur overwhelming overhead especially for highly dynamic wireless networks.\cite{EWC}. Thus, it is worth studying how to effectively optimize the resources in such a dynamic wireless environment.

Recently, transfer learning (TL)\cite{TL} has been widely employed to handle dynamic data in wireless resource allocation problems such as power control\cite{TL1} and beamformer design\cite{TL2}. However, once an NN model has adapted to the new environment by using TL, it would degrade or even overwrite the previously learned model, and thus the performance in the previous environment degrades significantly\cite{TL_disadvantage1, TL_disadvantage2}, which is termed as the catastrophic forgetting (CF) phenomenon. Besides, the performance of TL largely depends on the selection of the pre-trained model. Motivated by these challenges, we summarize the difficulties of dealing with the resource allocation problems in a dynamic wireless environment as: \textbf{How to achieve good performances under different CSI distributions without CF}.  

To achieve good performance under different CSI distributions, meta-learning\cite{meta, meta theo1} is a potential technique, where a good model initialization learned from a large amount of data with different distributions can help achieve good performance and fast adapt to new samples. The efficiency of meta-learning techniques in processing the new samples has been extensively studied in resource allocation problems\cite{TL2, meta_method1, meta_method2}. In \cite{TL2}, a downlink beamformer design based on meta-learning has been proposed to enable fast adaptation to a new testing wireless environment. In \cite{meta_method1}, the authors aimed to fast adapt to new network topology with limited data for the power control problem. Specifically, the ordinary black-box meta-learning technique has been improved by using the modular meta-learning, which can optimize a series of modules and quickly re-combine them when solving a new task. The authors in \cite{meta_method2} summarized the applications of meta-learning-based methods in wireless networks. However, the aforementioned works mainly focus on the improvement of fast adaption of meta-learning but the CF challenge is not considered.

As for the CF phenomenon, it can be potentially solved by the continual learning (CL)\cite{CL_survey1, CL_survey2}, which aims to incrementally learn new knowledge without forgetting previously learned knowledge. There have been a great number of works studying the CL and they can be roughly classified into three categories, i.e., regularization based methods\cite{EWC, SI}, dynamic NN architecture based methods\cite{CL1, CL2}, and memory box based methods\cite{CL3, CL4}. Among the aforementioned three categories, the first one is the most popular since the latter two would increase the training overhead due to the increase in the number of neurons or the size of the memory box. Specifically, the regularization based methods mainly study how to evaluate the importance of parameters and select less-important parameters to be modified in response to new data. This parameter evaluation and selection process is termed as selective plasticity in corresponding work. However, the design of selective plasticity in the aforementioned regularization based methods highly depend on the manual hyperparameter adjustment, which is impractical in practical applications. Therefore, it is necessary to apply the learning ability of NNs to achieve the goal of ‘learning to continually learn’. Inspired by the neuromodulatory processes of CL in human brain, there have been several papers on enabling the selective plasticity of NNs by using neuromodulation-based techniques \cite{brain1, brain2}, making the aforementioned goal possible.

In this paper, we take the classic sum-rate maximization (SRM) problem in the $K$-user interference network as example to study the dynamic DL technology. Specifically, we consider an episodically dynamic wireless environment, where the CSI distribution changes over periods and remains stationary within each period. Then, we develop a novel framework named meta-gating to overcome the aforementioned difficulties by achieving the following three important goals, where the former two are proposed for fulfilling good performance under different CSI distributions and the third goal is for overcoming the CF problem.
\begin{itemize}
    \item \textbf{Seamlessness}: The proposed method can achieve good sum-rate performance over all periods, which means that the sum-rate variance should be small enough so that it is unaware of changes in the CSI distribution.
    \item \textbf{Quickness}: The proposed method should well adapt to the new wireless environment with few training samples.
    \item \textbf{Continuity}: The proposed method can achieve good sum-rate performance in a new wireless environment without forgetting what has learned in previous environments/periods. Besides, it should not depend on the manual hyperparameter adjustment.
\end{itemize}

The proposed meta-gating framework consists of an inner network and an outer network. Specifically, for the former two goals, we propose a dual-loop training method by combining the model-agnostic meta-learning (MAML) algorithm with the unsupervised training. With such a design, the inner network is able to achieve good sum-rate performance on different channel distributions through a few number of stochastic gradient descent (SGD) iterations because of the suitable initialization. As for the goal of ‘continuity’, we adopt the regularization method and design an element-wise gating operation to multiply the outputs of the inner and outer networks, aiming to evaluate the importance of inner network’s parameters under different CSI distributions and then decide which subset of the inner network should be activated. Thus, it results in selective plasticity of the inner network by affecting its back propagation, where the selective plasticity is the core of regularization based methods in CL.

In summary, the main contributions of this work are highlighted as follows.
\begin{itemize}
    \item We propose a general framework to enable NNs to solve the resource allocation problems in a dynamic wireless environment, including the network architecture and the training method. The proposed framework can achieve three important goals, i.e., seamlessness, quickness, and continuity, to satisfy the requirements of a dynamic wireless environment via meta-learning and continual learning. 
    \item The proposed framework is model-agnostic, i.e, the inner and outer networks can be implemented as any NN models. Specifically, except that the number of outputs of the inner and outer networks need to be the same, the inner and outer networks have no other constraints, e.g., the kind of NNs, the number of neurons in the hidden layer, and the number of hidden layers.
    \item We provide rigorous analysis for the proposed framework in terms of the testing performance and generalization ability. Furthermore, in order to mathematically explain the CF problem, we propose a metric named channel distribution similarity (CDS) to measure the similarities between channels under different distributions.
\end{itemize}

The rest of the paper is organized as follows. The problem formulation and the meta-gating framework are given in Section \ref{system}. Section \ref{meta gating} introduces the comprehensive process of meta-gating framework for resource allocation problem. The theoretical analysis is introduced in Section \ref{theo}. Section \ref{simulation} presents the simulation results and performance analysis. Finally, this paper is concluded in Section \ref{conclusion}.
\section{Problem Formulation and the Meta-Gating Framework} 
\subsection{Problem Formulation}\label{system}

We consider an episodically dynamic wireless environment, where the CSI distribution changes over periods and remains constant within each period. Scenarios with the considered dynamic environment can be widely found in practice. For example, when a user drives from indoor to outdoor or moves from a highly dense place to an open place within a period of time, the CSI distribution will change accordingly (e.g., from Rayleigh fading with NLoS to Rician fading with LoS). Mathematically, we formulate the resource allocation problem in such a dynamic environment as follows
\begin{subequations}\label{dynamic_problem}
\begin{align}
\mathcal{P}_1:\quad\quad&\max_{\mathbf{p}(\mathbf{h})}\quad \mathbb{E}_{\mathbf{h}\sim M(\mathbf{h})}[Z(\mathbf{p}(\mathbf{h}), \mathbf{h})], \\
\mbox{s.t.}\quad
&\mathbf{J}(\mathbf{p}(\mathbf{h})) \leq 0,
\end{align}
\end{subequations}
where random variable $\mathbf{h}$ represents the instantaneous CSI (i.e., inputs of the NN-based models), $\mathbf{p}(\mathbf{h})$ denotes its corresponding instantaneous resource allocation strategy (i.e., outputs of the NN-based models), function $Z$ evaluates the instantaneous performance of strategy $\mathbf{p}(\mathbf{h})$, and $\mathbf{J}$ is a vector utility function to constrain the strategy $\mathbf{p}(\mathbf{h})$. Let $M(\mathbf{h})=\{m_1(\mathbf{h}), \cdots, m_t(\mathbf{h}), \cdots, m_T(\mathbf{h})\}$ represent the channel distributions in all periods, where $m_t(\mathbf{h})$ denotes the specific channel distribution in period $t$. 

Problem $\mathcal{P}_1$ aims to maximize the expectation of the evaluation function $Z(\cdot)$ to achieve good performance in an episodically dynamic wireless environment, i.e. find a strategy $\mathbf{p}(\mathbf{h})$ to maximize function $Z(\cdot)$ under constraints $\mathbf{J}$ in each period. 

\subsection{Overview of the Proposed Meta-Gating Framework} \label{meta gating}
In this subsection, we present the overall architecture of the proposed meta-gating framework and its training method for solving Problem $\mathcal{P}_1$. 

\subsubsection{Architecture of the Meta-Gating Framework}
\begin{figure}[htp]
    \centering
    \includegraphics[width=9cm]{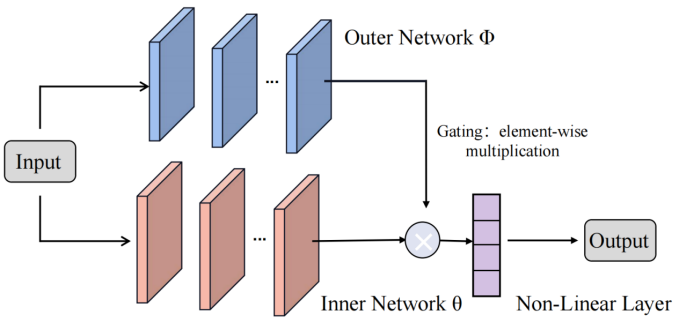}
    \vspace{-0.3cm}
    \caption{Architecture of meta-gating framework.}
    \label{fig: overall framework}
\end{figure}

As mentioned in Section \ref{introduction}, we attempt to utilize the learning ability of NNs to achieve the selective plasticity. Therefore, a dual-network structure is proposed, where the outer network extracts the characteristics of each CSI distribution. It aims to ensure the performance of the inner network under the previous CSI distribution when the inner network adapts to samples from a new CSI distribution. Specifically, as shown in Fig. \ref{fig: overall framework}, the proposed meta-gating network consists of an inner network, an outer network, and a non-linear layer, where both inner and outer networks are implemented as general NNs. Except that the number of outputs of the inner and outer networks need to be the same, there are no other constraints, e.g., the kind of NNs, the number of neurons in the hidden layer and the number of hidden layers. The inner and outer networks are connected through the gating operation, which refers to as element-wise multiplication of the output vectors of the inner and outer networks. After the multiplication, the results are input to the non-linear layer to obtain the final outputs.

\subsubsection{Training Procedure}

In this part, to achieve aforementioned three important goals, we design a training procedure for the proposed meta-gating framework, which is based on the model-agnostic meta-learning (MAML) algorithm\cite{MAML} and the unsupervised learning. 

\begin{figure}[htp]
    \centering
    \includegraphics[width=9cm]{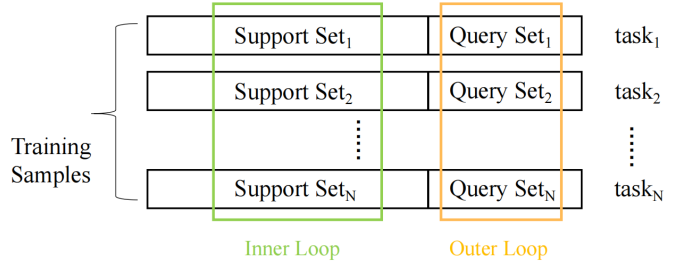}
    \vspace{-0.5cm}
    \caption{Dataset construction for the proposed framework.}
    \label{dataset}
\end{figure}

Different from the general DL where the wireless networks with different channel states can be directly used as different training samples, the training sample in the proposed training method refers to as a \textit{task}. Specifically, one task consists of a support set and a query set as shown in Fig. \ref{dataset}, both containing the samples in general DL. The samples in each support set and query set are randomly selected from the different channel distributions.

The proposed training procedure consists of an inner loop and an outer loop, where the inner loop is employed to update the inner network parameters $\bm{\theta}$ on the support set and the outer loop is for updating the outer network parameters $\bm{\phi}$ on the query set. Specifically, the parameters of the inner network are optimized by Adam optimizer\cite{adam} for $J$ iterations on the support set with the loss function $\mathcal{L}(\bm{\theta},\bm{\phi})$. During each of these $J$ forward propagation, the outputs of the inner network are gated, i.e. element-wisely multiplied, by the outputs of the outer network, which enables selective activation of the inner network by modifying its ultimate outputs during the forward propagation. Moreover, the gating to the inner network influences the update of the Adam optimizer and would result in selective plasticity during back propagation. After $J$ inner loop iterations, the parameters of the inner networks on task $i$ are denoted by ${\bm{\theta}}_J^i$, which will be used in the subsequent outer loop. As for the outer loop, the parameters of the outer network $\bm{\phi}$ are updated with the query sets and a meta loss $\mathcal{L}_{meta}$, which is calculated based on ${\bm{\theta}}_J^i$ and $\bm{\phi}$. The detailed training procedure of the proposed meta-gating framework is summarized in Algorithm \ref{A1}. 
\begin{table}[htp]
\vspace{-0.4cm}
	\setlength{\abovecaptionskip}{-2pt}
	\setlength{\belowcaptionskip}{-6pt}
	\begin{algorithm}[H]
		\caption{\textbf{Training Procedure of Meta-Gating Framework}}
		\label{A1}
		{\small
			\begin{algorithmic}[1]
			\State\textbf{Input:} training samples with size $N_m$, denoted as $\mathcal{T}=\{[S_1,Q_1],[S_2,Q_2],\ldots,[S_{N_m},Q_{N_m}]\}$, outer network learning rate $\alpha$, inner network learning rate $\beta$, batch size of training samples $B$.
			\State\textbf{Initialization:} parameters of the outer network $\bm{\phi}$, parameters of the inner network $\bm{\theta}$.
			\For{epoch=$1,2,\ldots$} \quad// \textit{Outer loop starts}
			\State Sample $B$ tasks from $\mathcal{T}$, let $\mathcal{L}_{\rm{meta}}=0$.
			\For{$i=1,2,\ldots, B$} 
			\For{$j=1,2,\ldots, J$} \quad// \textit{Inner loop starts}
			\State {$\bm{\theta}^i_j \longleftarrow \bm{\theta}^i_{j-1}-\beta\triangledown_{\bm{\theta}^i_{j-1}}\mathcal{L}(\bm{\phi},\bm{\theta}^i_{j-1}; S_i)$};
			\EndFor\quad // \textit{Inner loop ends}
			\State {$\mathcal{L}_{\rm{meta}} = \mathcal{L}_{\rm{meta}} + \mathcal{L}(\bm{\phi},\bm{\theta}^i_J; Q_i)$};
			\EndFor
			\State$\bm{\phi} \longleftarrow \bm{\phi}-\frac{\alpha}{B}\triangledown_{\bm{\phi}}\mathcal{L}_{\rm{meta}}$;
			\EndFor \quad// \textit{Outer loop ends}
		    \end{algorithmic}}
	\end{algorithm}
\end{table}

Following the aforementioned training procedure, the framework can well achieve aforementioned three goals and the reasons are as follows. First, the proposed framework can achieve the fast adaptation with small amount of samples because of the suitable initialization obtained by the MAML method. Thus, the proposed training method can well achieve the goal of ‘seamlessness’ and ‘quickness’. Secondly, the selective plasticity is achieved by the gating operation. Specifically, the importance of model parameters in response to different CSI distributions is different. The outer network is trained by the outer loop of Algorithm 1 with tasks from multiple CSI distributions. Therefore, the outer network can learn to evaluate the importance of inner network’s parameters under different CSI distributions, and then decide which subset of the inner network should be activated. By the gating operation, the meta-learned outer network can convey the decision to the inner network and thus indirectly influence the back propagation of the inner network. In this way, the inner network can perform well under both the previous and current CSI distribution, so as to overcome the CF problem.\footnote{Similarly, the graceful forgetting ability can be achieved by adjusting the learning abilities of inner network and outer networks, e.g, increasing the number of layers of the inner network within a certain range or adding a mask to the gating operation between the inner and outer network.} The testing procedure of the proposed framework is summarized in Algorithm \ref{A2}.

\begin{table}[htp]
\vspace{-0.5cm}
	\setlength{\abovecaptionskip}{-2pt}
	\setlength{\belowcaptionskip}{-6pt}
	\begin{algorithm}[H]
		\caption{\textbf{Testing Procedure of Meta-Gating Framework}}
		\label{A2}
		{\small
			\begin{algorithmic}[1]
			\State\textbf{Input:} sequential testing samples with size $N_m^{te}$, denoted as $\mathcal{T}^{te}=\{[S^{te}_1,Q^{te}_1],[S^{te}_2,Q^{te}_2],\ldots,[S^{te}_{N_m^{te}},Q^{te}_{N_m^{te}}]\}$, number of adaptation samples in $S^{te}_i$, denoted as $N_a$, inner-update learning rate $\beta$, {meta-learned parameters of the outer network, ${\bm{\phi}}^{*}$, and the inner network, ${\bm{\theta}}^{*}$}.
			\State Set $T_{train} = [\ ]$;
			\For{$i=1,2,\ldots, N_m^{te}$}
			\State $T_{train} = T_{train} + \mathcal{T}^{te}_i$;
			\State Randomly select $N_a$ samples from $S_i^{te}$ to form new $S_i^{te}$ for the following $J_q$ iterations;
			\For{$j=1,2,\ldots, J_q$} 
			\State $\bm{\theta}_j \longleftarrow \bm{\theta}_{j-1}-\beta\triangledown_{\bm{\theta}_{j-1}}\mathcal{L}({\bm{\phi}}^*,\bm{\theta}_{j-1}; S_i^{te})$;
			\EndFor
			\State Record $\mathcal{L}({\bm{\phi}}^*,\bm{\theta}_{J_{q}}; T_{train})$;
			\EndFor
		    \end{algorithmic}}
	\end{algorithm}
	\vspace{-0.5cm}
\end{table}

\section{Meta-Gating Framework for Resource Allocation Problem}
In this section, we take the SRM problem in a $K$-user interference network as an example to concretize $Z(\cdot)$, $\mathbf{J}(\cdot)$, and $\mathbf{p}(\mathbf{h})$ in Problem $\mathcal{P}_1$. Then, for the aforementioned example, two widely-used network models: GNN and CNN are adopted with the proposed meta-gating framework to further demonstrate its model-agnostic property.
\subsection{System Model of \textit{K}-User Interference Network}
\begin{figure}[bpth]
    \centering
    \includegraphics[width=6cm]{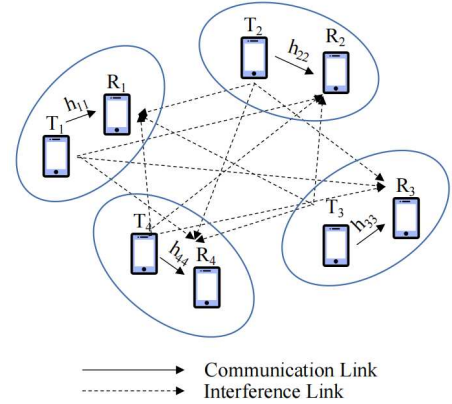}
    \caption{System model of the $K$-user interference network.}
    \label{fig: beam}
\end{figure}

As depicted in Fig. \ref{fig: beam}, there are $K$ transceiver pairs where each transmitter and receiver are equipped with $N_t$ and one antennas, respectively. It is assumed that transmissions on the $K$ transceiver pairs occur simultaneously using the same frequency band. Let $\mathbf{v}_k$ denote the beamformer of the $k$-th transmitter and $s_k$ denote the transmit signal. The received signal at receiver $k$ is $\mathbf{y}_k = \mathbf{h}^{H}_{kk}\mathbf{v}_ks_k +\sum^{K}_{j\neq k} \mathbf{h}^H_{jk}\mathbf{v}_js_j+n_k$, where $\mathbf{h}_{kk} \in \mathbb{C}^{N_t}$ denotes the direct channel vector between the $k$-th transceiver pair, $\mathbf{h}_{jk} \in \mathbb{C}^{N_t}$ denotes the interference channel vector from transmitter $j$ to receiver $k$, and $n_k \in \mathbb{C}$ denotes the additive noise following the complex Gaussian distribution $\mathcal{CN}(0,\sigma^2)$.\\
\indent Then, the signal-to-interference-plus-noise ratio (SINR) of receiver $k$ is expressed as
\begin{equation}
    \gamma_k = \frac{|\mathbf{h}^{H}_{kk}\mathbf{v}_k|^2}{\sum^{K}_{j\neq k}| \mathbf{h}^H_{jk}\mathbf{v}_j|^2+\sigma^2}.\label{origin SINR}
\end{equation}
\indent The optimization goal is to find the optimal beamformer matrix $\mathbf{V}=[\mathbf{v}_1,\cdots,\mathbf{v}_K]^T\in \mathbb{C}^{K\times N_t}$ to maximize the sum rate, i,e., $Z(\cdot)=\sum_{k=1}^{K} {\rm{log}}_2(1+\gamma_k)$. 

Finally, Problem $\mathcal{P}_1$ in this example can be concretized as
\begin{subequations} \label{beamforming problem}
\begin{align}
\mathcal{P}_2:\quad\quad&\max_{\mathbf{V}}\quad \mathbb{E}_{\mathbf{h}\sim M(\mathbf{h})}w_k\sum_{k=1}^{K} {\rm{log}}_2(1+\gamma_k),\\
\mbox{s.t.}\quad 
&{\Vert \mathbf{v}_k \Vert}_2^2\leq P_{{\rm{max}}}, \forall k,
\end{align}
\end{subequations}
where $w_k$ denotes the weight for the $k$-th transceiver pair and $P_{\rm{max}}$ represents the maximum transmit power of each transmitter.
\subsection{Meta-Gating GNN for Problem $\mathcal{P}_2$}
\subsubsection{Scenario Modeling}
In this part, we first model the wireless environment as a graph, and then formulate Problem $\mathcal{P}_2$ as a graph optimization problem.

In general, the wireless environment can be modeled as a weighted directed graph with both node and edge features. Formally, a graph can be represented as a four tuple $\mathcal{G} = (\mathcal{V}, \mathcal{E}, f, \bm{\alpha})$, where $\mathcal{V}$ is the set of nodes and $\mathcal{E}$ is the set of edges. For each node in $\mathcal{V}$, function $f$ maps it to its corresponding feature vector. For each edge in $\mathcal{E}$, it has a corresponding weight $\alpha(i, j)\in \bm{\alpha}$.
\begin{figure}[htp]
    \centering
    \subfigure[Graph modeling for the considered problem.]{
    \includegraphics[width=8.5cm]{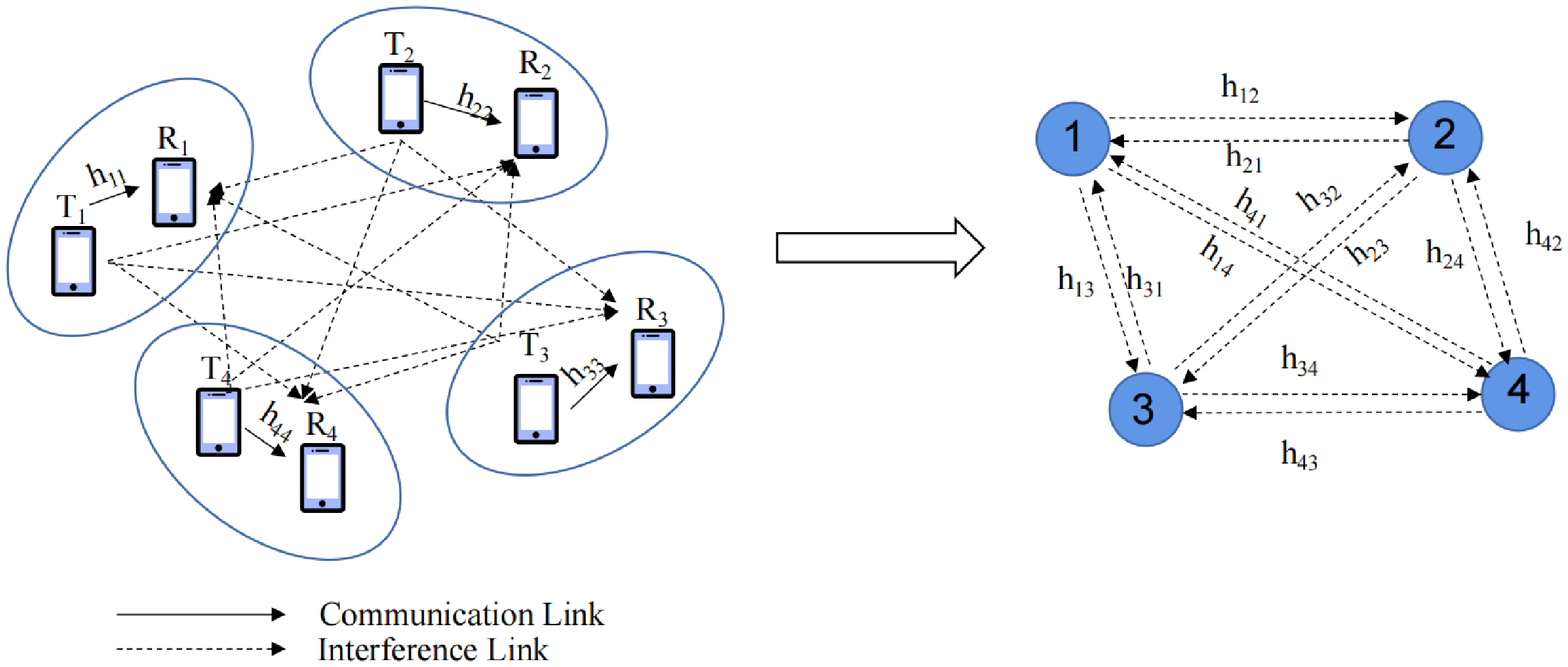}
    \vspace{-0.5cm}
    \label{fig: graph_model}}
    \subfigure[Important parts of the meta-gating GNN.]
    {
    \centering
    \includegraphics[width=7.5cm]{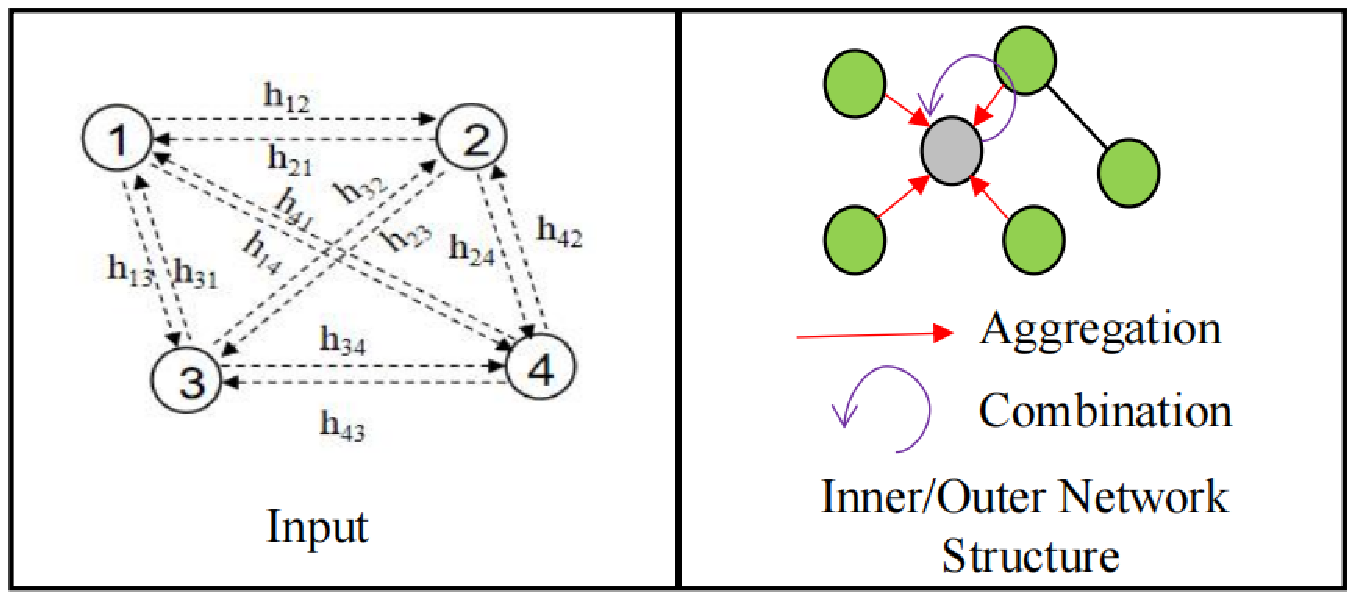}
    \label{fig: GNN_framework}
    }
    \caption{An illustration of meta-gating GNN for Problem $\mathcal{P}_2$.}
    \label{fig: meta-gating GNN}
\end{figure}

\indent Following the modeling in \cite{GNN2}, our considered system in Fig. \ref{fig: beam} can be modeled as a graph model in Fig. \ref{fig: graph_model}, where the $k$-th transceiver pair is treated as the $k$-th node in the graph. Moreover, the node feature matrix $\mathbf{Z}\in \mathbb{C}^{|\mathcal{V}|\times(N_t+2)}$ is given by $\mathbf{Z}_{(k,:)} = [\mathbf{h}_{kk}, w_k, \sigma^2]^T$, and the weight matrix $\bm{\alpha}$ is given by
\begin{equation}
    \bm{\alpha}_{(j,k)} = 
     \left\{
\begin{aligned}
&\mathbf{0},\quad(j,k)\notin \mathcal{E},\\
&\mathbf{h}_{jk}, {\rm{otherwise}},\\
\end{aligned}
\right.
\end{equation}
where vector $\mathbf{0}$ is a zero vector with a size of $N_t$.

Then the SINR in (\ref{origin SINR}) can be rewritten with the notations $\mathbf{Z}$, $\bm{\alpha}$, and $\mathbf{V}$ as follows
\begin{equation}
    \gamma_k = \frac{|\mathbf{Z}^H_{(k,1:N_t)}\mathbf{v}_k|^2}{\sum_{j\neq k}^K|\bm{\alpha}_{(j,k)}\mathbf{v}_j|^2+\mathbf{Z}_{(k,N_t+2)}}.
\end{equation}
Finally, Problem $\mathcal{P}_2$ in each period can be reformulated as 
\begin{subequations} \label{final problem}
\begin{align}
&\max_{\mathbf{V}}\quad \sum_{k=1}^{K}\mathbf{Z}_{(k,N_t+1)}{\rm{log}}_2(1+\gamma_k),\\
\mbox{s.t.}\quad 
&{\Vert \mathbf{v}_k \Vert}_2^2\leq P_{{\rm{max}}}, \forall k.
\end{align}
\end{subequations}
\subsubsection{Forward Propagation}

As depicted in Fig. \ref{fig: GNN_framework}, the input data of meta-gating GNN is the graph model in Fig. \ref{fig: graph_model} and the final outputs are the optimal beamformer matrix $\mathbf{V}$ in each period. Both inner and outer networks are implemented as wireless communication graph convolution network (WCGCN) \cite{GNN2}, which belongs to the message passing graph neural network (MPGNN). Before introducing the WCGCN model, we first describe the mechanism of MPGNN. Specifically, the update process (key operation of GNNs) of the $n$-th layer at node $k$ in an MPGNN is describe as 
\begin{equation}
    \bm{x}_k^n = \gamma^n(\bm{x}_k^{n-1},\beta^n_{j\in\mathcal{N}(k)}([\bm{x}_k^{n-1}, \bm{e}_{jk}])),
\end{equation}
where $\bm{x}_k^n$ represents the hidden state of the $n$-th layer at node $k$, $\bm{x}_k^0$ is the input node feature vector of node $k$, $\mathcal{N}(k)$ denotes the neighbors of node $k$, and $\bm{e}_{jk}$ is the input edge feature vector of edge $(j,k)$. Moreover, $\beta(\cdot)$ is the function that aggregates information from the neighbors of node $k$ and $\gamma(\cdot)$ is the function that combines the aggregated information with its own information, which can be seen in Fig. \ref{fig: GNN_framework}. Furthermore, $\beta(\cdot)$ can be further simplified by applying NNs as follows
\begin{equation}
    \beta(\bm{x}) = \psi(f(\bm{x})),
\end{equation}
where $\psi$ is implemented by some simple functions, such as ${\rm{MAX}}$ and ${\rm{SUM}}$, and $f$ is the existing NN structure.

In the WCGCN model, ${\rm{MAX}}$ is utilized as function $\psi$, and two different multi-layer perceptrons (MLPs) are applied to function $\gamma(\cdot)$ and $f(\cdot)$, respectively. Thus, its update process of node $k$ can be expressed as $\bm{x}_k^n = {\rm{MLP}}_{2}(\bm{x}_k^{n-1},{\rm{MAX}}_{j\in\mathcal{N}(k)}\left\{{\rm{MLP}}_1([\bm{x}_j^{n-1}, \bm{e}_{jk}])\right\})$. Then, the forward propagation of node $k$ in the proposed meta-gating GNN can be expressed as
\begin{align}
    \mathbf{x}^n_k &= {\rm{MLP_2}}\left(\mathbf{x}_k^{n-1}, {\rm{MAX}}_{j\in\mathcal{N}(k)}\left\{{\rm{MLP_1}}\left([\mathbf{x}_j^{n-1}, \bm{\alpha}_{(j,k)}]\right)\right\}\right),\\
    \hat{\mathbf{x}}^n_k &= {\rm{MLP_4}}\left(\hat{\mathbf{x}}_k^{n-1}, {\rm{MAX}}_{j\in\mathcal{N}(k)}\left\{{\rm{MLP_3}}\left([ \hat{\mathbf{x}}_j^{n-1}, \bm{\alpha}_{(j,k)}]\right)\right\}\right),\\
    \mathbf{y}_k &= \sigma\left(\mathbf{x}^N_k\odot\hat{\mathbf{x}}^M_k\right),
\end{align}
where $\bm{x}_k^0 = \hat{\bm{x}}_k^0 = \bm{Z}_{(k,:)}$ is the input node feature, $\bm{\alpha}_{(j,k)}$ is the weight matrix that represents the input edge feature, $N$ and $M$ denote the number of layers in the inner and outer networks, respectively, $\sigma(\cdot)$ is a differentiable normalization function in the non-linear layer, and $\odot$ denotes the element-wise multiplication operation.
\subsubsection{Back Propagation}

Due to the lack of an optimal solution to Problem $\mathcal{P}_2$, unsupervised training that directly maximizes the sum rate is applied for solving the considered problem. The loss function to be minimized can be written as

\begin{equation} \label{beam}
\resizebox{.95\hsize}{!}{$
    \mathcal{L}(\bm{\theta},\bm{\phi}) = -\sum_{k=1}^K\mathbf{Z}_{(k,N_t+1)}{\rm{log}}_2\left(1+\frac{|\mathbf{Z}^H_{(k,1:N_t)}\mathbf{v}_k(\bm{\theta}, \bm{\phi})|^2}{\sum_{j\neq k}^K|\bm{\alpha}_{(j,k)}\mathbf{v}_j(\bm{\theta}, \bm{\phi})|^2+\mathbf{Z}_{(k,N_t+2)}}\right).$}
\end{equation}

\subsubsection{Complexity Analysis}
For the meta-gating GNN, the inner and outer networks are implemented by WCGCNs. The complexity of WCGCN is $\mathcal{O}(L(|E|+|V|))$, where $L$ is the number of layers of WCGCN, $|E|$ denotes the size of edge set, and $|V|$ denotes the size of node set. Therefore, the complexity of the proposed framework is $\mathcal{O}({\rm{MAX}}\{N, M\}(|\mathcal{E}|+|\mathcal{V}|))$.
\subsection{Meta-Gating CNN for Problem $\mathcal{P}_2$}
\subsubsection{Scenario Modeling}
\begin{figure}[htp]
    \centering
    \subfigure[Picture-like pixel modeling for the considered problem.]{
    \includegraphics[width=8.5cm]{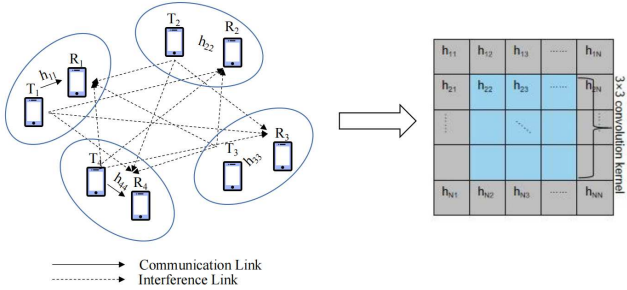}
    \vspace{-0.5cm}
    \label{fig: CNN_model}}
    \subfigure[Important parts of the meta-gating CNN.]
    {
    \centering
    \includegraphics[width=7.5cm]{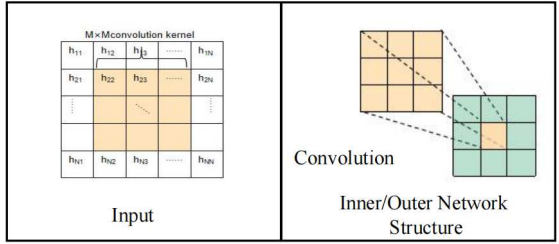}
    \label{fig: CNN_framework}
    }
    \caption{An illustration of meta-gating CNN for Problem $\mathcal{P}_2$.}
    \label{fig: meta-gating CNN}
\end{figure}

CNN has been widely used in DL, e.g., it can extract spatial features from an image for classification. In a wireless environment, CNN is generally utilized to exploit the spatial features in channel state. It is because that the nearby receiver plays more significant role in determining the beamformer of the $k$-th transmitter. Besides, CNN has fewer number of trainable parameters compared to DNN and can greatly reduce the training overhead. Based on this, we model $\mathbf{H}=\{\mathbf{h}_{j,k}, \forall j,k\}$ in Problem $\mathcal{P}_2$ as a picture-like pixel structure, as shown in Fig. \ref{fig: CNN_model}.
\subsubsection{Forward Propagation}

As depicted in Fig. \ref{fig: CNN_framework}, the input data of meta-gating CNN is the picture-like pixel structure and the final outputs are the the optimal beamformer matrix $\mathbf{V}$ in each period. Both inner and outer networks are implemented as general CNNs\cite{CNN}. The forward propagation in the proposed framework can be expressed as 
\begin{align}
    \mathbf{x}^n &= {\rm{ReLU}}\left({\rm{Conv}}(\mathbf{x}^{n-1};c_{\rm{in}},c_{\rm{out}}, s)\right), \notag\\
	  \mathbf{u}^N &= {\rm{FC}}\left({\rm{MP}}(\mathbf{x}^N\right), \quad 1\leq n \leq N,\\
    \hat{\mathbf{x}}^m &= {\rm{ReLU}}\left({\rm{Conv}}(\hat{\mathbf{x}}^{m-1};\hat{c}_{\rm{in}},\hat{c}_{\rm{out}}, \hat{s})\right), \notag\\ \hat{\mathbf{u}}^M &= {\rm{FC}}\left({\rm{MP}}(\hat{\mathbf{x}}^M\right), \quad 1\leq m \leq M,\\
    \mathbf{y} &= \sigma\left(\mathbf{u}^N\hat{\mathbf{u}}^M\right),
\end{align}
where $\mathbf{x}^0=\hat{\mathbf{x}}^0=\mathbf{H}$, $\mathbf{x}^n$ and $\hat{\mathbf{x}}^m$ denote the $n$-th and $m$-th hidden state of the inner network and the outer network, respectively. $\mathbf{u}^N$ and $\mathbf{u}^M$ represent the outputs of the inner and outer networks, respectively. ${\rm{ReLU}}$ represents the rectified linear unit layer to prevent the negative values, ${\rm{FC}}$ represents the fully-connected layer, and ${\rm{MP}}$ denotes the max-pooling operation. $\sigma(\cdot)$ is a differentiable normalization function, and $\mathbf{y}$ denotes the final outputs. $N$ and $M$ represent the layer number of the inner and outer networks, respectively, and ${\rm{Conv}}$ represents the convolution layer that performs two-dimensional spatial convolution of the input data. The size of the convolution layer is denoted as $s(\hat{s})$ and its depth is set to $c_{\rm{in}}, c_{\rm{out}}(\hat{c}_{\rm{in}}, \hat{c}_{\rm{out}})$. 
\subsubsection{Back Propagation}
Similarly, we employ the unsupervised training for the considered problem and the loss function to be minimized can be written as
\begin{equation} \label{powercontrol}
\resizebox{.9\hsize}{!}{$
    \mathcal{L}(\bm{\theta},\bm{\phi}) = -\sum_{k=1}^Kw_k{\rm{log}}_2\left(1+\frac{|\mathbf{h}^{H}_{kk}\mathbf{v}_k(\bm{\theta},\bm{\phi})|^2}{\sum^{K}_{j\neq k}| \mathbf{h}^H_{jk}\mathbf{v}_j(\bm{\theta},\bm{\phi})|^2+\sigma^2}\right)$}.
\end{equation}
\subsubsection{Complexity Analysis}
For the meta-gating CNN, the inner and outer networks are implemented by CNNs. The complexity of CNN is $\mathcal{O}(\sum\limits_{l=1}^{L}Q_l^2S_l^2C_{l-1}C_l)$, where $L$ is the number of layers of CNN, $Q_l$ denotes the output size of $l$-th layer, $S_l$ represents the size of convolution kernel, and $C_l$ is the number of channels in the $l$-th layer. Therefore, the complexity of the proposed framework is $\mathcal{O}({\rm{MAX}}\{\sum\limits_{n=1}^{N}Q_n^2s_n^2c_{n-1}c_n, \sum\limits_{m=1}^{M}{\hat{Q}_m}^2{\hat{s}_m}^2\hat{c}_{m-1}\hat{c}_m\})$, where $Q$ and $\hat{Q}$ denote the output size of inner and outer networks, respectively.
\section{Theoretical Analysis of the Meta-Gating Framework}\label{theo}
In this section, we theoretically analyze the performance of the proposed meta-gating framework. Specifically, we first propose a metric named CDS to measure the distances between different channel distributions, which is employed to explain the CF phenomenon between different channel distributions in the following simulation part. Then, we analyze the impact of the number of update round, $J_q$, to demonstrate that the value of $J_q$ cannot be chosen too large, which exactly satisfies the requirement of fast adaptation. Finally, we analyze the generalization ability of the proposed framework in terms of the gradient of its loss function with respect to the trained parameters.

\subsection{Distances Between Different Channel Distributions}
\begin{figure}[htp]
    \centering
    \includegraphics[width=8.7cm]{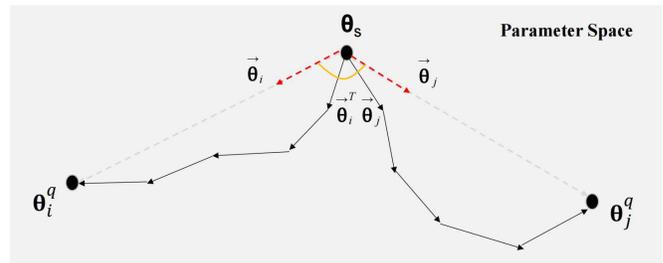}
    \vspace{-0.3cm}
    \caption{Visualizations of the update of parameters to different channel distributions. The black arrows represent the SGD on the support set of the corresponding channel distribution, the dotted red arrows represent the trajectory direction vector and the orange arc represents the inner product between the two adaptation trajectories.}
    \label{TL theory}
\end{figure}

In this part, CDS is designed to measure the difference between channel distributions from parameter space. It is because that the outputs of NN-based model on different input data distributions may not be quite different, even if the distances between input data distributions are large. Therefore, the influence of different data distributions on the outputs of NN-based model cannot be judged only from the input data space. Based on the above analysis, we don’t need to obtain the absolute value of the distances between different channel distributions. Instead, we need to measure the impact of different distributions on the output of NN-based model with the same initialization. In the following, we present the detailed mechanism of CDS. Specifically, we assume that a pre-trained model $f$ adapts to a channel distribution $\mathcal{C}_i$ starting from $\bm{\theta}_s$ and moves to the final solution $\bm{\theta}_i^{q}$ by performing $q$ SGD iterations steps. Then, the parameters' adaptive trajectory to channel distribution $\mathcal{C}_i$ starting from $\bm{\theta}_s$ is defined as the sequence of iterations, which is denoted as $\{\bm{\theta}_s, \bm{\theta}_i^1, \bm{\theta}_i^2, \cdots, \bm{\theta}_i^{q}\}$. To alleviate the challenges in dealing with trajectories of multiple steps in a parameter space of a very high dimension, the trajectory direction vector $\vec{\bm{\theta}}_i$ can be defined as
\begin{equation}
\vec{\bm{\theta}}_i \triangleq \frac{\bm{\theta}_i^{q}-\bm{\theta}_s}{\Vert \bm{\theta}_i^{q}-\bm{\theta}_s \Vert_2}.   
\end{equation}
Fig. \ref{TL theory} presents the SGD update trajectories for the pre-trained model to adapt to channel distribution $\mathcal{C}_i$ and channel distribution $\mathcal{C}_j$, respectively. Based on the above analysis, CDS is finally defined as the inner product between their direction vectors.
\begin{equation}
CDS = \vec{\bm{\theta}}_i^T\vec{\bm{\theta}}_j.   
\end{equation}

Compared with the KL-divergence that only measures the distances between different distributions from the input data space, the proposed metric measures from the model parameters
space. It takes the characteristics of the model into consideration so that the impact of different data distributions on the output of NN-based model can be well judged.

\subsection{Impact of the Number of Update Round——Fast Adaptation}\label{impact_jq}

In this part, we focus on the impact of the number of update round, $J_q$, on the performance of the proposed meta-gating framework.

We denote the dataset of channel $c\sim \mathcal{C}$ as $D_c$, where the number of testing samples is denoted as $N_m^{te}$. Assume that we run $J_q$ gradient descent steps on $D_c$ to obtain the updated model $\bm{\theta}_c^{J_q} = \bm{\theta}^* - \beta[\triangledown\mathcal{L}_{D_c}(\bm{\theta}^*|\bm{\phi}^*)+\sum_{t=1}^{J_q-1}\triangledown\mathcal{L}_{D_c}(\bm{\theta}_c^t|\bm{\phi}^*)]$ for channel $c$. Let $\bm{\theta}^*$ and $\bm{\phi}^*$ denote the initializations of inner and outer networks, respectively, and both are learned from the proposed training method. $\bm{\theta}_c^*$ represents the optimal model parameters of the channel $c\sim \mathcal{C}$. As a premise, we first introduce some necessary definitions.
\begin{Definition}
  \textbf{Lipschitz continuity and smoothness}\footnote{This assumption is widely used in the analysis of deep learning, such as \cite{LP1, LP2}}.
  \emph{Function $g(\theta)$ is $G$-Lipschitz continuous if $\Vert g(\theta_1)-g(\theta_2)\Vert_2\leq G\Vert\theta_1-\theta_2\Vert_2$ with a constant $G$. And $g(\theta)$ is called $L$-smooth if $\Vert \triangledown g(\theta_1)-\triangledown g(\theta_2)\Vert_2\leq L\Vert\theta_1-\theta_2\Vert_2$ with a constant $L$}.
\end{Definition}

\begin{Definition}\label{excess risk}
  \textbf{Excess Risk}.
  \emph{ER$(\bm{\theta}_c^{J_q})=\mathbb{E}_{c\sim \mathcal{C}}\mathbb{E}_{D_c}[\mathcal{L}(\bm{\theta}^{J_q}_c,\bm{\phi}^*)-\mathcal{L}(\bm{\theta}^*_c,\bm{\phi}^*)]$, where $\mathcal{L}(\cdot)$ denotes the expected loss on $\bm{\theta}_c$}. 
\end{Definition}

It evaluates the loss difference between $\bm{\theta}_c^{J_q}$ and the optimal model $\bm{\theta}^*_c$ on all samples $D_c$ with all channels $c\sim \mathcal{C}$, and a smaller value means a better $\bm{\theta}_c^{J_q}$. In the following, excess risk is used to analyze the influence of the number of update round $J_q$, i.e., the testing performance of fast adaption to new samples.

\begin{Theo}\label{theo 1}
(\textbf{Testing Performance Analysis}). 
\emph{Suppose that the loss function $\mathcal{L}$ is $G$-Lipschitz continuous and $L$-smooth w.r.t. both the inner and outer network parameters ($\bm{\theta}$ and $\bm{\phi}$). Assume that $\beta$ obeys $\beta\leq\frac{1}{L}$ and denote $\rho=1+2\beta L$. Then for any $c\sim\mathcal{C}$ and $D_c$ with size $N_m^{te}$, we have}
{\small
\begin{align}
    ER(\bm{\theta}_c^{J_q})
    &\leq \underbrace{\frac{2G^2(\rho^{J_q}-1)}{N_m^{te}L}}_{\mathcal{O}(\frac{\rho^{J_q}}{N_m^{te}})}+\mathbb{E}_{c\sim \mathcal{C}}\mathbb{E}_{D_c}[\mathcal{L}_{D_c}(\bm{\theta}^{J_q}_c,\bm{\phi}^*)-\mathcal{L}(\bm{\theta}^*_c,\bm{\phi}^*)].\label{theo1}
\end{align}
}
\end{Theo}

The detailed proof can be found in Appendix \ref{appendix_a}. Theorem \ref{theo 1} demonstrates that the excess risk ER$(\bm{\theta}_c^{J_q})$ of the channel-specific updated model $\bm{\theta}_c^{J_q}$ for channel $c$ is mainly determined by three key factors, i.e., the testing sample number $N_m^{te}$ (more precisely, the size of the support set in the testing samples), the value of $J_q$, and the expected loss $\mathbb{E}_{c\sim \mathcal{C}}\mathbb{E}_{D_c}[\mathcal{L}_{D_c}(\bm{\theta}^{J_q}_c,\bm{\phi}^*)-\mathcal{L}(\bm{\theta}^*_c,\bm{\phi}^*)]$ between the adapted parameter $\bm{\theta}_c^{J_q}$ and the optimal model $\bm{\theta}^*_c$. Note that, a larger $N_m^{te}$ leads to a smaller upper bound for the first term in (\ref{theo1}). However, in order to reduce the overhead, the amount of online update data should not be too large, thus, $N_m^{te}$ cannot be too large. Besides, an intuitive way to reduce the excess risk is to increase the value of $J_q$ which however increases the upper bound of the first term. It is because that $\rho$ is usually slightly larger than $1$ given a small learning rate $\beta$. Therefore, to make a fair trade-off between the first and second terms in (\ref{theo1}), $J_q$ should not be large, which accords with the impact of the number round $J_q$ in the following simulation parts of Section \ref{simulation}.

\subsection{First-Order Optimality Analysis——Generalization Ability}

To measure the testing performance of the adapted parameters $\bm{\theta}_c^{J_q}$ in terms of first-order optimality, we first introduce the expected population gradient.
\begin{Definition}\label{Expected Population Gradient}
  \textbf{Expected Population Gradient}.
  \emph{Let EPG$(\bm{\theta}_c^{J_q})=\mathbb{E}_{c\sim \mathcal{C}}\left[\Vert\mathbb{E}_{D_c}[\triangledown \mathcal{L}(\bm{\theta}_c^{J_q}|\bm{\phi}^*)]\Vert_2^2\right]$ denote the gradient of the loss function $\mathcal{L}(\bm{\theta}_c, \bm{\phi}^*)$ on all samples $D_c$ and all channels $c\sim \mathcal{C}$}.
\end{Definition}

 In the following, we will apply this expected population gradient as the metric to measure the testing performance of $\bm{\theta}_c^{J_q}$ so as to verify the generalization ability of learned initializations $\bm{\theta}^*$.

\begin{Theo}\label{theo 2}
(\textbf{First-order Optimality Analysis}). 
\emph{Suppose that the loss function $\mathcal{L}$ is $G$-Lipschitz continuous and $L$-smooth w.r.t. both the inner and outer network parameters ($\bm{\theta}$ and $\bm{\phi}$). Assume that $\beta$ obeys $\beta\leq\frac{1}{L}$ and denote $\rho=1+2\beta L$. Then for any $c\sim\mathcal{C}$ and $D_c$ with size $N_m^{te}$, we have}
{\small
\begin{equation}
    EPG(\bm{\theta}_c^{J_q})\leq\frac{8G^2(\rho^{J_q}-1)^2}{{N_m^{te}}^2}+2\mathbb{E}_{c\sim\mathcal{C}}\mathbb{E}_{D_c}\left[\Vert\triangledown\mathcal{L}_{D_c}(\bm{\theta}_c^{J_q}|\bm{\phi}^*)\Vert_2^2\right].\label{epg}
\end{equation}
}
\end{Theo}

The detailed proof can be found in Appendix \ref{appendix_b}. Theorem \ref{theo 2} reveals the importance of the empirical gradient $\mathbb{E}_{D_c}\left[\Vert\triangledown\mathcal{L}_{D_c}(\bm{\theta}_c^{J_q}|\bm{\phi}^*)\Vert_2^2\right]$ on determining the expected population gradient EPG$(\bm{\theta}_c^{J_q})$. Specifically, when the learned initializations $\bm{\theta}^*$ are close to the first-order stationary points of the empirical risk $\mathcal{L}_{D_c}(\bm{\theta}_c, \bm{\phi}^*)$, a small value of $J_q$ (a few gradient descent steps) can already guarantee a very small gradient $\triangledown\mathcal{L}_{D_c}(\bm{\theta}_c^{J_q}|\bm{\phi}^*)$ of the adapted parameters $\bm{\theta}_c^{J_q}$. Besides, the value of $J_q$ is small and the testing samples are usually sufficient, which has been proved in Theorem \ref{theo 1}. Therefore, the first term of (\ref{epg}) is also small. Finally, the proposed framework is proved to have a good generalization ability because of the small value of EPG$(\bm{\theta}_c^{J_q})$.\\

\section{Simulation Results}\label{simulation}

In this section, we conduct simulations to demonstrate the effectiveness of the proposed meta-gating framework. All codes are implemented in Python 3.9 with Pytorch 1.8.0 and we consider the following benchmarks for comparison, where the channel state samples refer to the samples from one specific CSI distribution.

\begin{itemize}
    \item \textbf{Joint} (Joint Training): It updates the model using all channel state samples.
    \item \textbf{Mismatch}: It trains the model under one of the channel state samples.
    \item \textbf{TL} (Transfer Learning): It trains the pre-trained model only using the current channel state samples, where the pre-trained model was trained under a given channel distribution.
    \item \textbf{EWC} (Elastic Weight Consolidation): It adds a penalty term to the loss function so as to prevent large changes in those parameters that are important to previous samples. The importance of the parameters is judged by the Fisher information matrix\cite{EWC}.
    \item \textbf{WoGate}: It updates model via traditional MAML method, i.e., without gating operation in the proposed framework.
\end{itemize}

\subsection{Simulation Results on Meta-Gating GNN} \label{beamforming_sim}
We consider $K$ transceiver pairs within an $R \times R$ area, where the transmitters are generated uniformly in the aforementioned area and the receivers are generated uniformly within $[d_{\rm{{min}}}, d_{\rm{{max}}}]$ from their corresponding transmitters. We adopt the channel model in \cite{channel_GNN} as follows
\begin{equation}
    h_{j,k} = L\left(\underbrace{\sqrt{\frac{\epsilon}{\epsilon+1}}\bm{\alpha}_t(\beta_t)\bm{\alpha}_r(\beta_r)^H}_{Los}+\underbrace{\sqrt{\frac{1}{\epsilon+1}}\hat{h}_{j,k}}_{NLos}\right),
\end{equation}
where $L$ denotes the large-scale fading including the path loss and shadowing, $\beta_t$ and $\beta_r$ denote the transmit and receive directions, respectively. We adopt the large-scale fading model in \cite{largescale_fading} and generate the following three standard types of channel distributions as the sequential input data.
\begin{itemize}
    \item \textbf{Channel 1}: $\epsilon=0$ and each channel state $\hat{h}_{jk}$ is generated according to a standard normal distribution, i.e.,
    \begin{equation}
    {\rm{Re}}(\hat{h}_{jk})\sim\frac{\mathcal{N}(0,1)}{\sqrt{2}},\quad {\rm{Im}}(\hat{h}_{jk})\sim\frac{\mathcal{N}(0,1)}{\sqrt{2}}, \forall j,k.
    \end{equation}
    \item \textbf{Channel 2}: $\epsilon=3$ dB, both $\beta_t$ and $\beta_r$ are uniformly generated from $[0, 2\pi]$, and each channel state $\hat{h}_{jk}$ is generated according to the Gaussian distribution with $0$ dB $K$-factor, i.e.,
    \begin{equation}
    {\rm{Re}}(\hat{h}_{jk})\sim\frac{1+\mathcal{N}(0,1)}{2},\quad {\rm{Im}}(\hat{h}_{jk})\sim\frac{1+\mathcal{N}(0,1)}{2}, \forall j,k.
    \end{equation}
    \item \textbf{Channel 3}: $\epsilon=0$, the shadow fading in $L$ is set as normal distribution with a standard deviation of $8$ dB, and each channel state $\hat{h}_{jk}$ is generated the same as \textbf{Channel 1}.
\end{itemize}

We generate $N_m=600$ \emph{tasks} as the training samples for Algorithm 1, where the support set and the query set in each task are composed by $2$ and $15$ channel state samples, respectively. Note that the channel state samples in each support set and query set are randomly selected from the three aforementioned channels. It is noticed that $600$ \emph{tasks} here are equivalent to $(2+15)\times 600=10,200$ channel state samples in general DL, which is sufficient to obtain a good model. As for the testing stage, $500$ channel state samples are generated for each channel, where we randomly split $20\%$ of these samples into support set for inner network's fine-tune process, and the rest $80\%$ into query set. Besides, we set the batch size of training samples $B$ and the number of adaptation samples $N_a$ as $5$ and $2$, respectively. Furthermore, we adopt the Adam optimizer with a learning rate of $0.0001$ to optimize the outer network and a learning rate of $0.001$ to optimize the inner network in the training stage. The Adam optimizer with a learning rate of $0.001$ is adopted to fine-tune the inner network in the testing stage.

Finally, in order to compare the sum-rate performance under different channel distributions, we normalize the sum rate by the weighted minimum mean-square error (WMMSE) algorithm \cite{wmmse}, which is termed as the ‘Normalized Sumrate’ in the following simulation results. The WMMSE algorithm is a classic optimization-based algorithm for sum-rate maximization in the $K$-user interference network and is usually used as an upper bound for such problems. In this section, we run WMMSE for $100$ iterations with the random initialization and take this value for normalization. The system parameters and network parameters are summarized in Table \ref{table1} and Table \ref{table2}, respectively.
\begin{table}[htp]
    \renewcommand\arraystretch{1}
	\small
	\caption{System Parameters}
	\label{table1}
	\centering
	\scalebox{1}{
	\begin{tabular}{c|c}
		\toprule
		\textbf{Parameter} & \textbf{Value}\\
		\hline\hline
		Transceiver pairs, $K$ & $10$\\
		\hline
		Area length, $R$ & $1,000$ m\\
		\hline
		\# of Transmitter antennas, $N_t$ & $8$\\
		\hline
		Transceiver pairs distance, $d_{\rm{min}}$, $d_{\rm{max}}$ & $2$ m, $65$ m\\
		\hline
		Noise power, $\sigma^2$& $-10$ dB\\
		\hline
		Maximum transmit power, $P_{\rm{max}}$& $1$ w\\
		\hline
		Weight for the $k$-th transceiver pair, $w_k$ & 1\\ 
		\bottomrule
	\end{tabular}}
\end{table}
\vspace{-0.3cm}
\begin{table}[htp]
    \renewcommand\arraystretch{1}
	\small
	\caption{Neural Network Parameters}
	\label{table2}
	\centering
	\scalebox{1}{
	\begin{tabular}{c|c}
		\toprule
		\textbf{Parameter} & \textbf{Value}\\
		\hline\hline
		Type of NN & WCGCN\cite{GNN2}\\
		\hline
		\makecell{Number of layers in outer and \\inner networks} & $3$, $2$\\
		\hline
		MLPs in outer/inner network & \makecell{\{$6N_t$, $64$, $64$\}, \\\{$64+4N_t$, $32$, $2N_t$\}}\\
		\hline
		Nonlinear function, $\sigma(x)$ & $\sigma(x)=\frac{x}{{\rm{max}}(\lVert x\rVert_2,1)}$\\
		\bottomrule
	\end{tabular}}
\end{table}
\subsubsection{Three Important Goals}
\begin{figure}[htp]
    \centering
    \begin{minipage}[t]{0.48\textwidth}
    \centering
    \includegraphics[width=8cm]{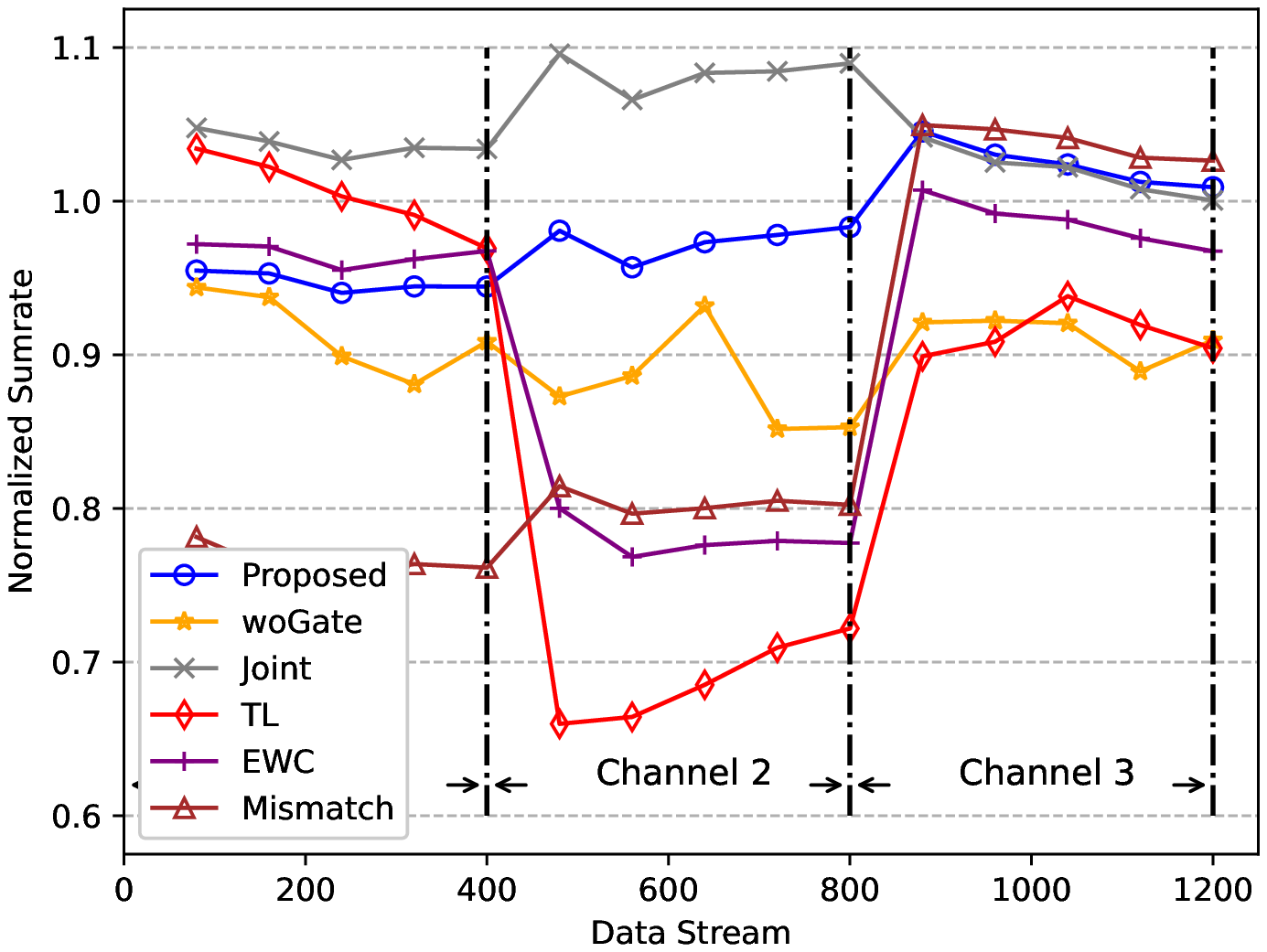}
    \vspace{-0.5cm}
    \caption{Sum-rate performances under different methods.}
    \label{fig: sumrate}
    \end{minipage}
    \begin{minipage}[t]{0.48\textwidth}
    \centering
    \includegraphics[width=8cm]{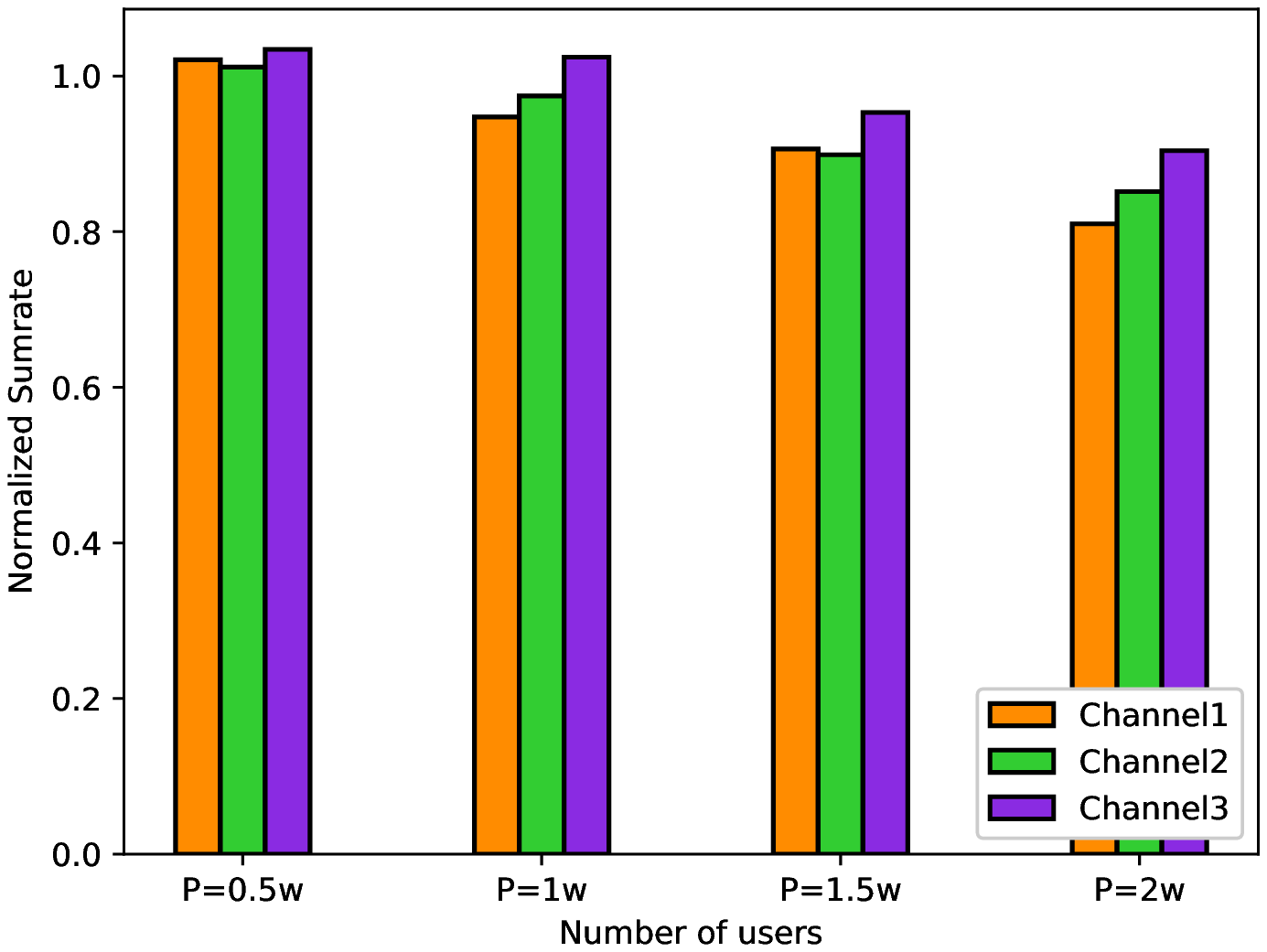}
    \vspace{-0.5cm}
    \caption{Sum-rate performance with different values of $P_{\rm{max}}$.}
    \label{fig: sumrate_diff_p}
    \end{minipage}
\end{figure}

(\romannumeral1) \textbf{Seamlessness}. Fig. \ref{fig: sumrate} compares the proposed meta-gating framework with the above benchmark algorithms on the sum-rate performance, where the channel state samples in ‘Mismatch’ are collected from ${\rm{channel}}_3$. Obviously, applying a model trained under one distribution to test the samples on another distribution does lead to a quite large sum-rate performance loss. Moreover, we can observe that the sum-rate performance of our meta-gating framework is better than that of TL under the premise of the same number of fine-tune samples, update rounds, and optimizer. It is mainly because that the proposed framework has better initializations compared to the TL and thus achieves better sum-rate performance using only a small number of update rounds. Besides, compared with TL, the MAML algorithm can obtain a suitable model initialization for all three channels. Thus, the average sum-rate performance is better than that of TL. 

Similar with the TL, the sum-rate performance of ‘WoGate’ on ${\rm{channel}}_2$
and ${\rm{channel}}_3$ are affected by the online update according to the previous channel state samples, which can be seen from the gap between ‘proposed’ and ‘woGate’. In the proposed framework, the inner network updates according to the current channel sample data, where only part of the model parameters that are selected by the meta-learned outer network will be updated. Therefore, the proposed framework can still achieve good sum-rate performance on the current channel state
sample and is not largely affected by the previous channel state samples. 

Table \ref{table3} presents the variance of the sum-rate performance among different channel distributions under different methods. It can be observed that the proposed framework has the smallest variance, indicating that the sum-rate performance on different channel distributions is basically similar. Therefore, the proposed framework can well achieve the goals of ‘seamlessness’.

\begin{table}[htp]
    \renewcommand\arraystretch{1}
	\small
	\caption{Variances of the normalized sum-rate performances under different methods.}
	\label{table3}
	\centering
	\scalebox{0.96}{
	\begin{tabular}{c| c| c}
		\toprule
		\textbf{Method} & (\textbf{Channel 1}, \textbf{Channel 2}) &(\textbf{Channel 2}, \textbf{Channel 3})\\
		\hline
		TL&$2.49\times 10^{-2}$&$1.27\times 10^{-2}$\\
		\hline
		EWC&$8.58\times 10^{-3}$&$1.05\times 10^{-2}$\\
		\hline
		Joint&$5.63\times 10^{-4}$&$1.04\times 10^{-3}$\\
		\hline
		WoGate&$3.04\times 10^{-4}$&$2.79\times 10^{-4}$\\
		\hline
		Proposed&$1.83\times 10^{-4}$&$6.20\times 10^{-4}$\\
		\bottomrule
	\end{tabular}}
\end{table}

Furthermore, we compare the sum-rate performance of the proposed meta-gating framework under different values of maximum transmit power $P_{\rm{max}}$, and the results are depicted in Fig. \ref{fig: sumrate_diff_p}. From the figure, the proposed framework achieves good sum-rate performance under different $P_{\rm{max}}$, i.e., the sum-rate performance is basically the same as the WMMSE algorithm. Moreover, Table \ref{table4} presents the variances of the normalized sum rate among different channel distributions under different values of $P_{\rm{max}}$. From the table, these variances are all quite small which further highlights the advantage of the proposed meta-gating framework in terms of ‘seamlessness’.
\begin{figure}[htp]
    \centering
    \begin{minipage}[t]{0.48\textwidth}
    \centering
    \includegraphics[width=8cm]{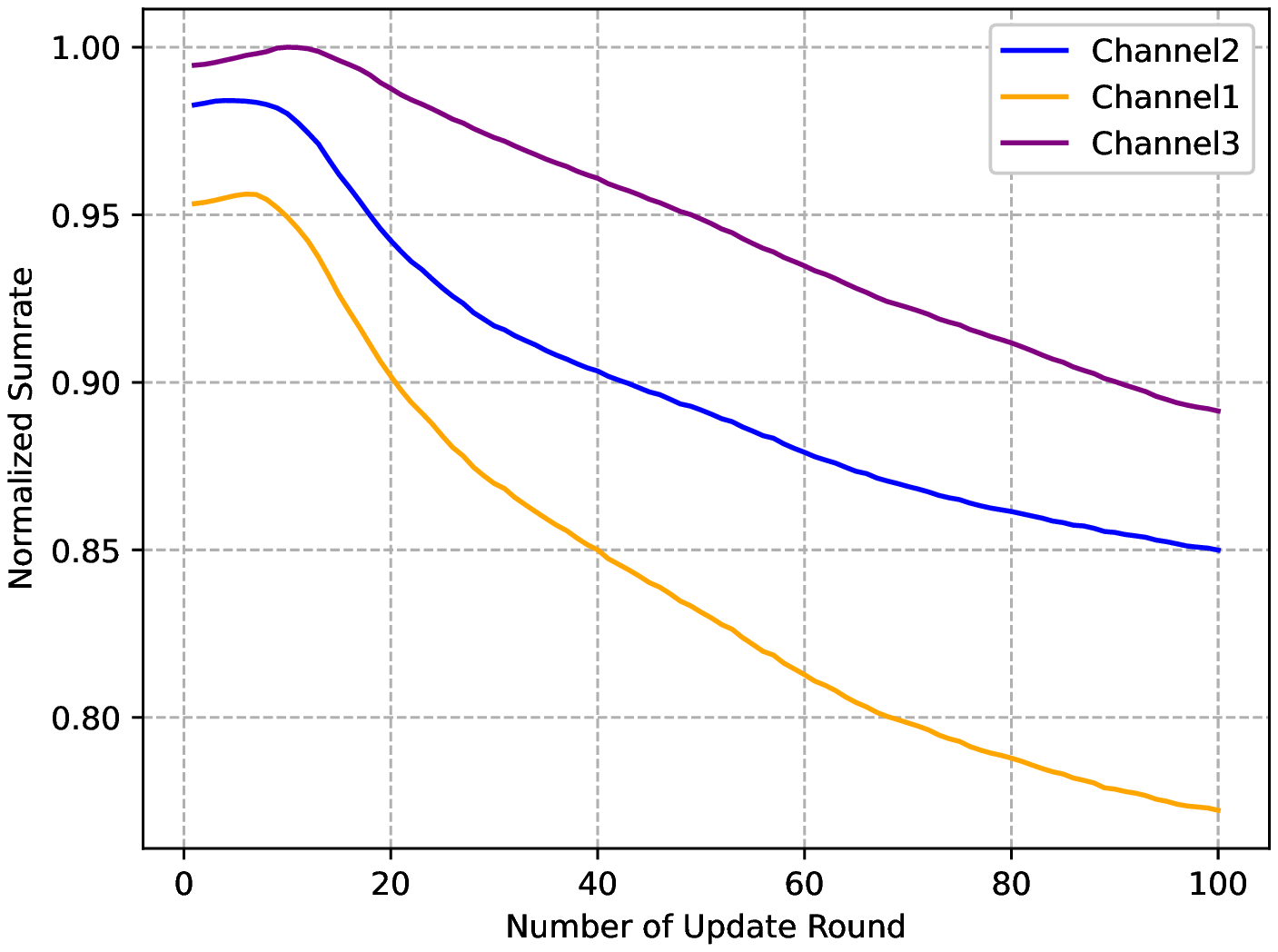}
    \vspace{-0.5cm}
    \caption{Sum-rate performance with different values of update round $J_q$.}
    \vspace{0.4cm}
    \label{fig: sumrate_diff_q}
    \end{minipage}
    \begin{minipage}[t]{0.48\textwidth}
    \centering
    \includegraphics[width=9cm,height=7cm]{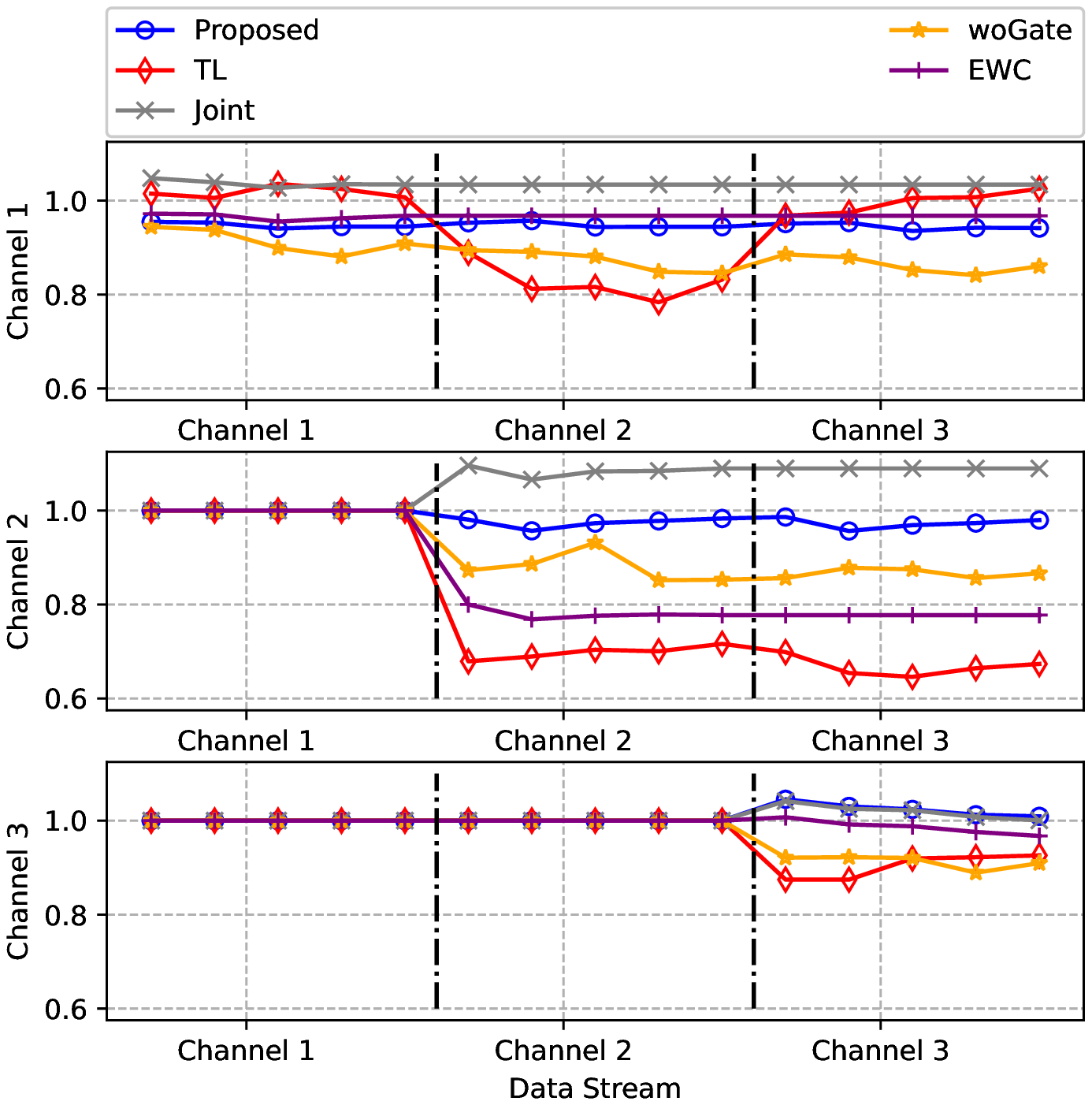}
    \vspace{-0.5cm}
    \caption{Capability for continuous adaptation of different methods.}
    \label{fig: notforgetting}
    \vspace{-0.3cm}
    \end{minipage}
\end{figure}

\begin{table}[htp]
    \renewcommand\arraystretch{1.2}
	\small
	\caption{Variances of the normalized sum-rate performances under different values of $P_{\rm{max}}$.}
	\label{table4}
	\centering
	\scalebox{0.85}{
	\begin{tabular}{c c c c c}
		\toprule
		$P_{{\rm{max}}}$ (W) & $0.5$&$1$&$1.5$&$2$\\
		\hline
		Variance&$8.69\times 10^{-5}$&$1.01\times 10^{-3}$&$5.78\times 10^{-4}$&$1.48\times 10^{-3}$\\
		\bottomrule
	\end{tabular}}
\end{table}

(\romannumeral2) \textbf{Quickness}. Fig. \ref{fig: sumrate_diff_q} depicts the impact of different values of $J_q$ on the sum-rate performance. From the figure, the proposed framework only needs a small value of $J_q$ to achieve good sum-rate performance under each channel distribution ($J_q=10$ in this simulation scenario). It is mainly because that the proposed meta-gating framework has good model initializations via the proposed training procedure. Besides, we can see that the sum-rate performance under different channel distributions first increases and then decreases with the increase of $J_q$. The degradation of the sum-rate performance is mainly caused by the severe overfitting on the small amount of adaptation samples $N_a$. In fact, the fine-tune process with small amount of samples exactly achieves the goal of ‘quickness’, where the value of $J_q$ is small to avoid serious overfitting phenomenon. The simulation results are consistent with the theoretical analysis in Section \ref{impact_jq}.

(\romannumeral3) \textbf{Continuity}. Fig. \ref{fig: notforgetting} depicts the ‘continuity’ capability of different methods. Specifically, it shows the sum-rate performance of the proposed framework on ${\rm{channel}}_j$ (the vertical axis) after updating according to the ${\rm{channel}}_i$ (the horizontal axis). In order to clearly illustrate the capability for continuous adaptation of different methods, we set the value of the normalized sum rate in ${\rm{channel}}_j$ as $1$ when $j \textgreater i$. From the figure, the sum-rate performance of TL on the previous channel suffers from a significant degradation when it adapts to the following new channel distribution. It is mainly because that the model in TL is only fine-tuned on the latest new samples. After learning the knowledge of new samples, the knowledge from the previous model may be altered or even overwritten, which thus results in significant performance deterioration on the previous samples. Similar results can be seen from ‘WoGate’. On the other hand, the proposed meta-gating framework utilizes the outer network to evaluate the importance of inner network’s parameters under different CSI distributions and then decide which subset of the inner network should be activated through the gating operation. Therefore, it can ensure the capability for continuous adaptation.

\begin{table}[htp]
    \renewcommand\arraystretch{1.3}
	\small
	\caption{Distance between each channel}
	\label{table5}
	\centering
	\scalebox{1}{
	\begin{tabular}{c c c c}
		\toprule
		Channel pair &(1,2)&(2,3)&(3,1)\\
		\hline
		CDS&$0.0099$&$0.0114$&$0.0177$\\
		\bottomrule
	\end{tabular}}
\end{table}

According to the analysis in Section \ref{theo}, we compute the similarity between the considered three channel distributions and the results are given in Table \ref{table5}. From the table, the distance between ${\rm{channel}}_1$ and ${\rm{channel}}_2$ is the largest, indicating that the distribution between these two channels are quite different. Therefore, it will cause the largest performance loss in ${\rm{channel}}_1$ when the model is updated according to the samples in ${\rm{channel}}_2$ in TL. Moreover, the CDS metric shown in Table \ref{table5} can explain the capability for continuous adaptation of the EWC method in Fig. \ref{fig: notforgetting}. Specifically, channel samples under each distribution are sequentially input during the EWC training stage. In order not to forget the knowledge learned on ${\rm{channel}}_1$, the update of model's parameters on ${\rm{channel}}_2$ will be affected by the consolation operation, where the distance between ${\rm{channel}}_1$ and ${\rm{channel}}_2$ is the largest. Therefore, it finally leads to poor sum-rate performance on ${\rm{channel}}_2$.

\begin{figure}[htp]
\centering
\begin{minipage}[t]{0.48\textwidth}
\centering
\includegraphics[width=8cm]{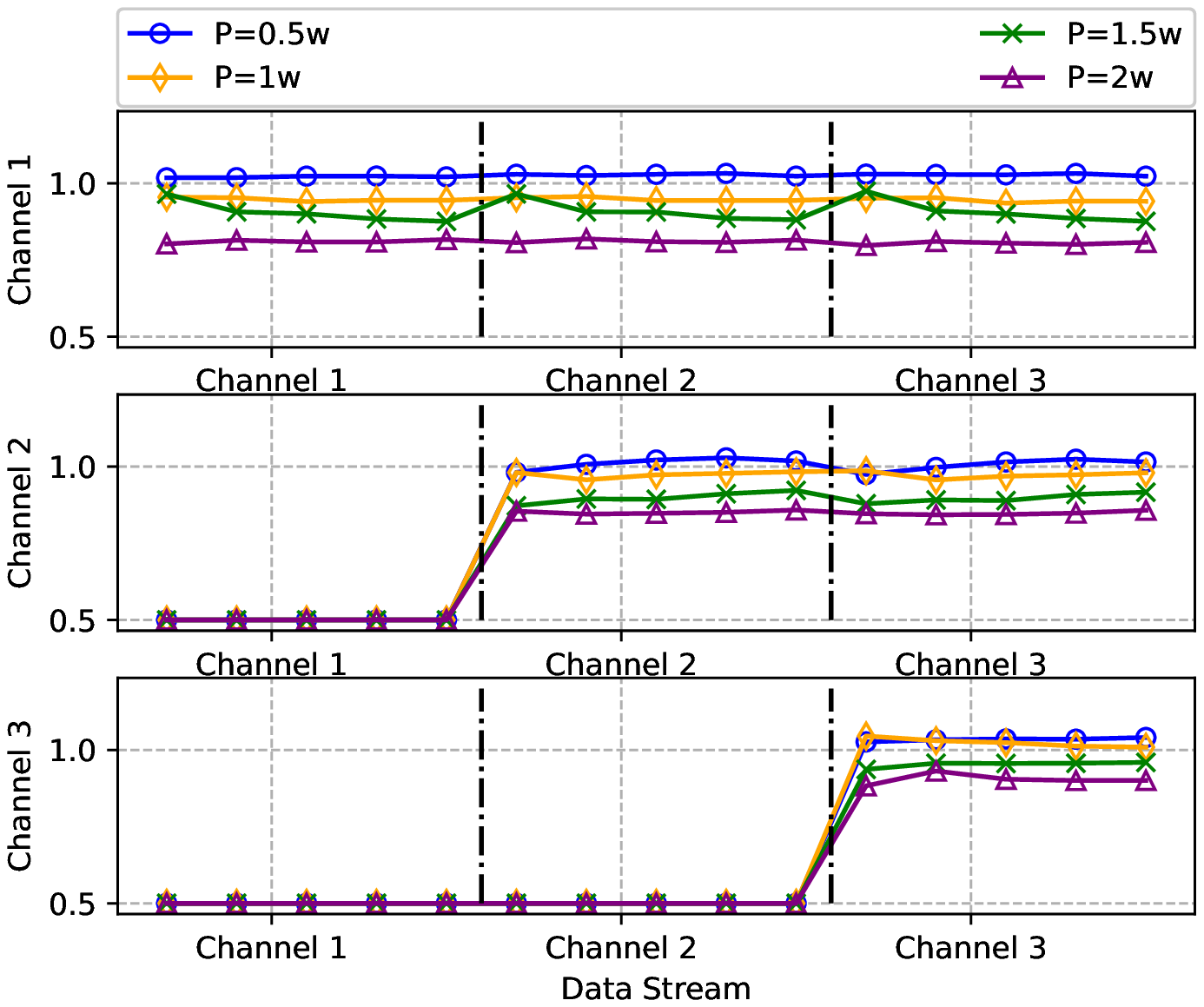}
\vspace{-0.3cm}
\caption{Capability for continuous adaptation with different values of $P_{\rm{max}}$.}
\label{fig: notforgetting_diff_p}
\end{minipage}
\begin{minipage}[t]{0.48\textwidth}
\centering
\includegraphics[width=8cm]{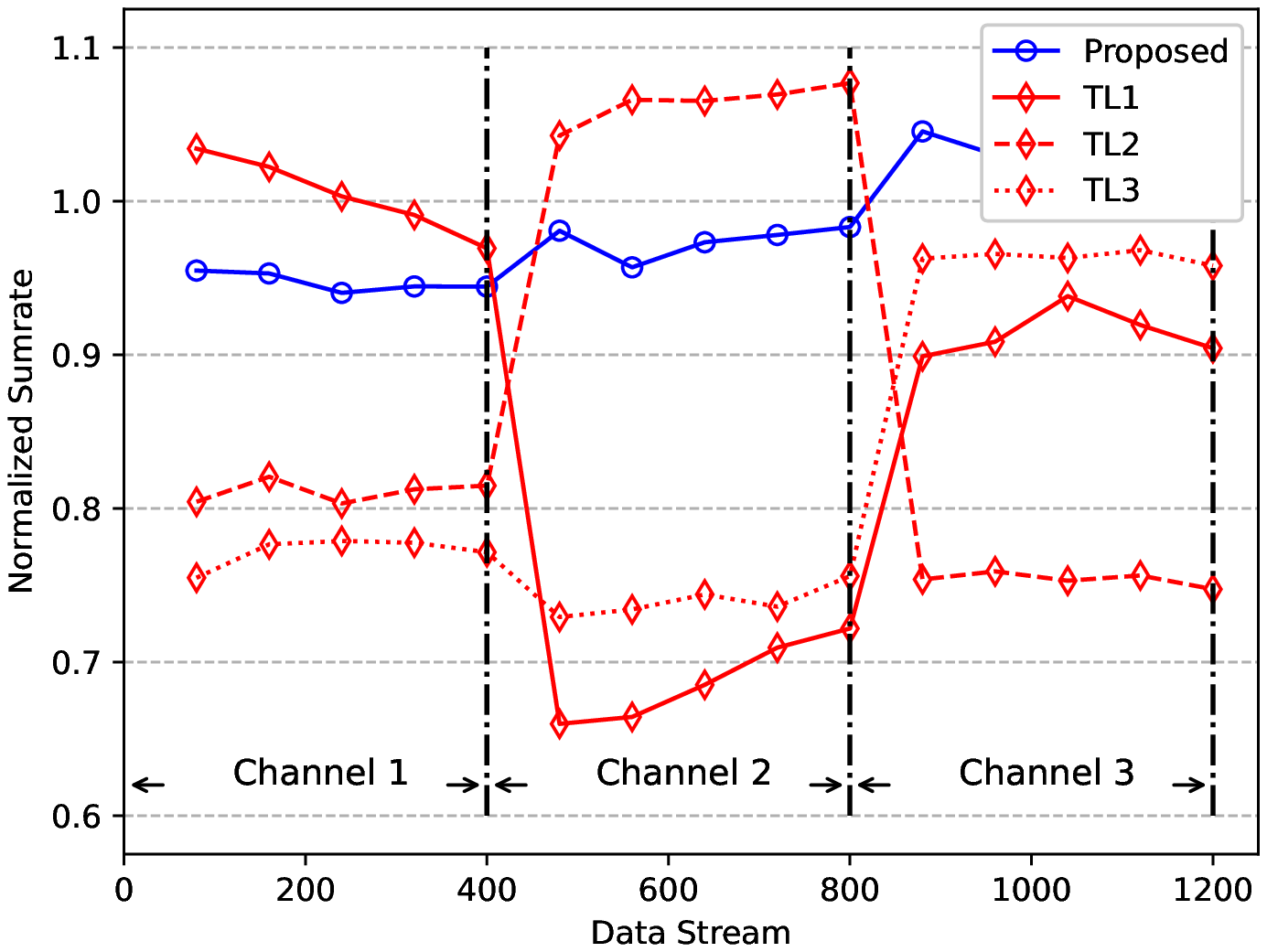}
\vspace{-0.3cm}
\caption{Effect of pre-train models on sum-rate performance of TL.}
\label{fig: sumrate_TL}
\end{minipage}
\vspace{-0.3cm}
\end{figure}

Furthermore, Fig. \ref{fig: notforgetting_diff_p} compares the capability for continuous adaptation of the proposed meta-gating framework under different values of $P_{\rm{max}}$. From the figure, the proposed framework achieves the good capability for continuous adaptation under each value of $P_{\rm{max}}$, i.e., the sum-rate performance on ${\rm{channel}}_i$ does not largely degrade when the model is updated according to the samples of the ${\rm{channel}}_j$. It further indicates the advantage of the proposed framework in terms of ‘continuity’.

\subsubsection{Performance Comparison with TL}
It is quite important for TL to select a suitable pre-train model since the model initializations make great influence on the adaptation. Fig. \ref{fig: sumrate_TL} presents the impact of different pre-train models on the sum-rate performance, where TL1, TL2, and TL3 represent the models trained with the samples of ${\rm{channel}}_1$, ${\rm{channel}}_2$, and ${\rm{channel}}_3$ as pre-train models, respectively. Since there does exist differences between each channel distribution, the sum-rate performance with different pre-train models will be quite different. In contrast, the proposed meta-gating framework achieves a better sum-rate performance because of its adaptivity on different channel distributions. 
\begin{figure}[htp]
\vspace{-0.5cm}
\centering  
\subfigure[Sum-rate performance.]{
\label{fig: EWC_sum}
\includegraphics[width=8.2cm]{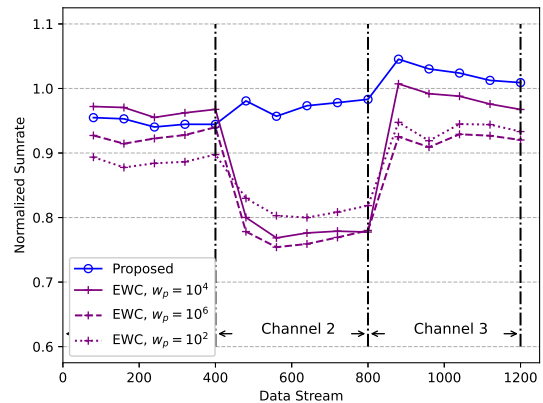}}
\subfigure[Capability for continuous adaptation.]{
\label{fig: EWC_CL}
\includegraphics[width=7.2cm]{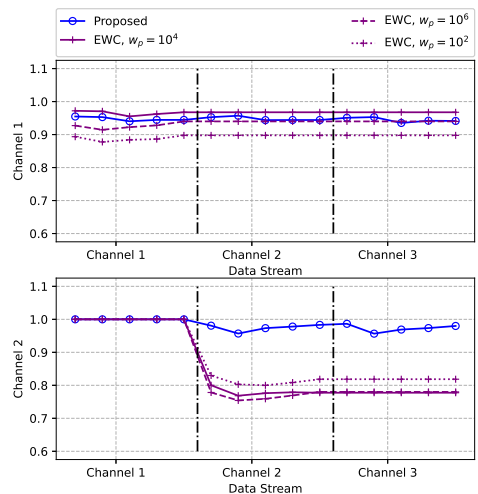}}
\vspace{-0.2cm}
\caption{The impact of $w_p$ on the performance of EWC method.}
\label{fig: EWC_sum+CL}
\vspace{-0.3cm}
\end{figure}
\subsubsection{Performance Comparison with EWC}
The performance of the EWC method largely depends on the coefficient of the penalty term, $w_p$. To verify it, we test the sum-rate performance on each channel distribution and the capability for continuous adaptation with $w_p= 10^2, 10^4, 10^6$, which can be seen in Fig. \ref{fig: EWC_sum+CL}. It is observed that a large value of $w_p$ can indeed improve the sum-rate performance on the previous channel, i.e., enhance the capability for continuous adaptation of the model, but with the cost of significant sum-rate performance loss on the current channel. Therefore, it is important for the EWC method to select an appropriate value of $w_p$, which is the disadvantage of the EWC method. Different from the EWC method that needs to be manually implemented, the proposed meta-gating framework can continuously achieve the good sum-rate performance because of the proposed training procedure and the gating operation.

\subsubsection{Scalability}
\begin{figure}[htbp]
\centering
\vspace{-0.5cm}
\begin{minipage}[t]{0.48\textwidth}
\centering
\includegraphics[width=8cm, height=6cm]{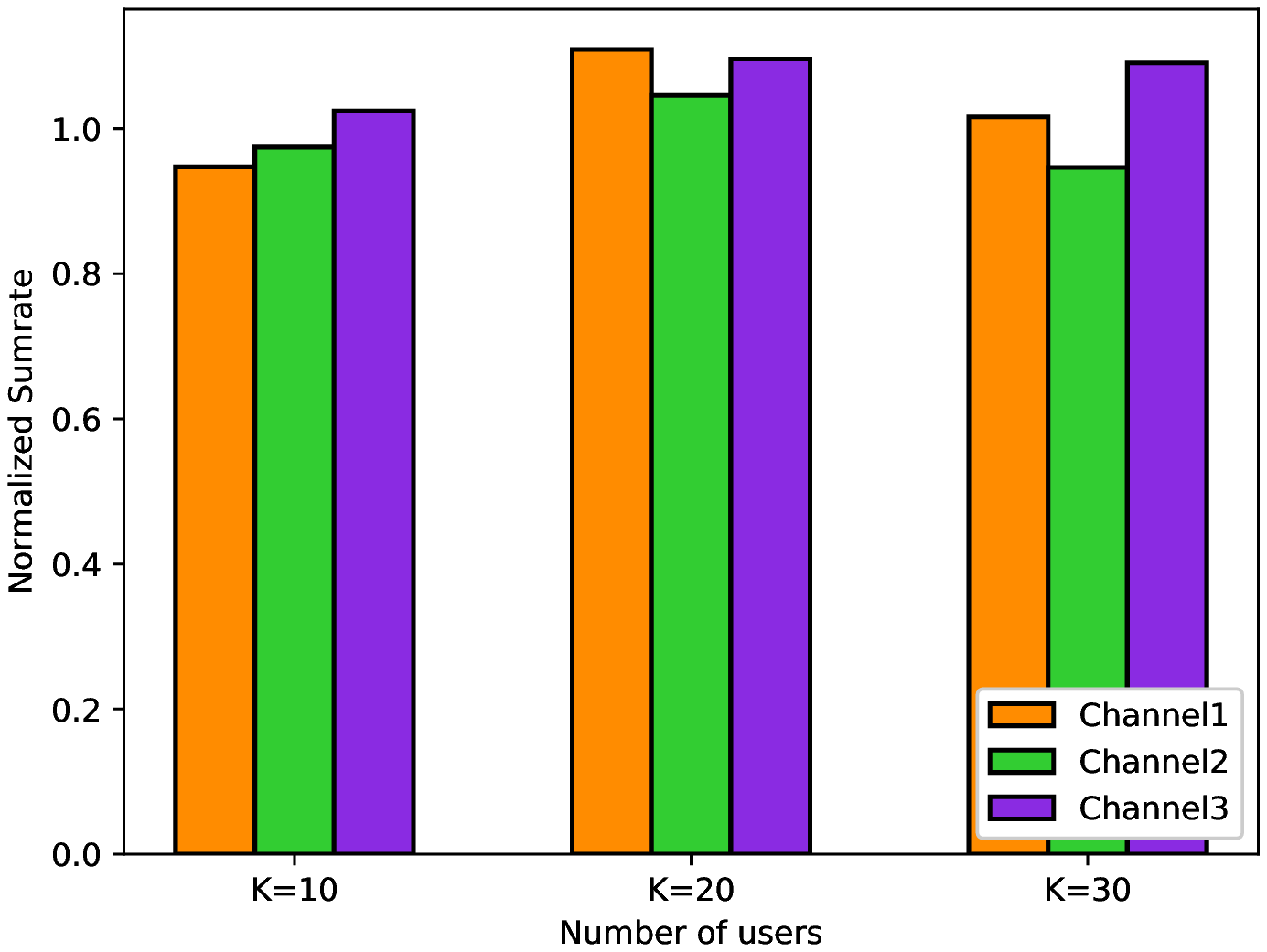}
\vspace{-0.3cm}
\caption{Sum-rate performances with different numbers of users.}
\label{fig: gen_sumrate}
\vspace{0.5cm}
\end{minipage}
\begin{minipage}[t]{0.48\textwidth}
\centering
\includegraphics[width=8cm]{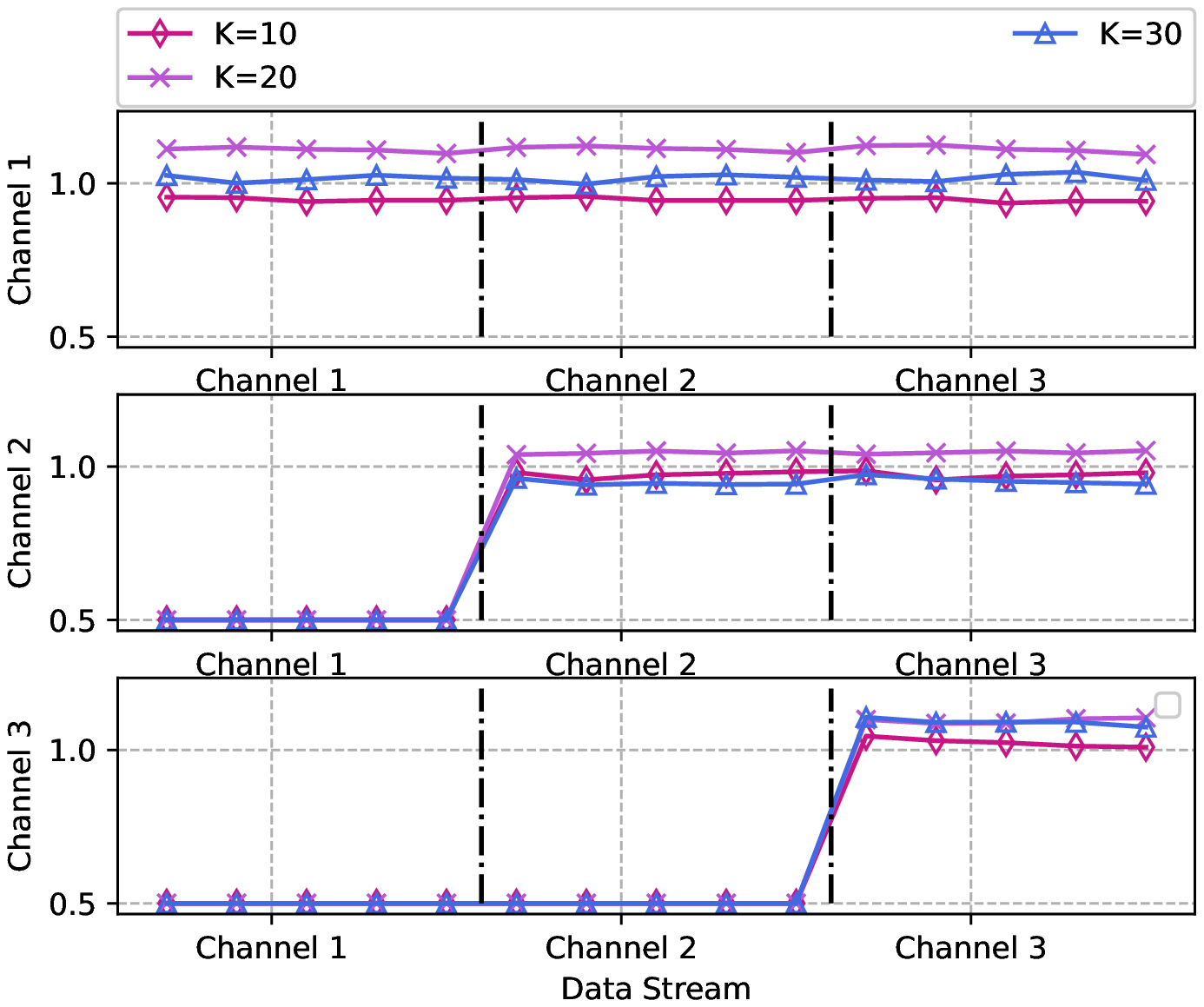}
\vspace{-0.3cm}
\caption{Capability for continuous adaptation with different numbers of users.}
\label{fig: gen_continuous}
\end{minipage}
\end{figure}

In this part, we test the sum-rate performance and the capability for continuous adaptation with different numbers of users ($K = 10$, $20$ and $30$) in the same area with radius $R = 1000$ m, where the number of update round $J_q$ is chosen as $10$. As depicted in Fig. \ref{fig: gen_sumrate} and Table \ref{table12}, the proposed meta-gating framework achieves good sum-rate performances with different numbers of users and the variances of the normalized sum rate among different channel distributions under different numbers of users are all quite small, which further highlights the advantage of the proposed meta-gating framework in terms of ‘seamlessness’. Besides, as shown in Fig. \ref{fig: gen_continuous},  the proposed framework achieves the good capability for continuous adaptation with different numbers of users, i.e., the sum-rate performance on ${\rm{channel}}_j$ does not largely degrade when the model is updated according to the samples of the ${\rm{channel}}_i$. These results demonstrate the good scalability of the proposed framework with more practical simulation scenario settings.

\begin{table}[htp]
    \renewcommand\arraystretch{1.2}
	\small
	\caption{Variances of the normalized sum-rate performances with different numbers of users.}
	\label{table12}
	\centering
	\scalebox{1}{
	\begin{tabular}{c c c c}
		\toprule
		$K$ & $10$&$20$&$30$\\
		\hline
		Variance&$1.01\times 10^{-3}$&$7.45\times 10^{-4}$&$3.47\times 10^{-3}$\\
		\bottomrule
	\end{tabular}}
\end{table}

\subsection{Simulation Results on Meta-Gating CNN}\label{powercontrol_sim}
In this part, we present the performance of meta-gating CNN on Problem $\mathcal{P}_2$ to demonstrate that the proposed framework is model-agnostic. Since the main purpose of this simulation part is to verify the model-agnostic nature of the proposed framework, we only consider a special case of Problem $\mathcal{P}_2$ with $N_t = 1$. Then, three standard types of random channels following the settings in \cite{continuous} are denoted as ${\rm{channel}}_1$, ${\rm{channel}}_2$ and ${\rm{channel}}_3$. Specifically, ${\rm{channel}}_1$ follows the Rayleigh fading, ${\rm{channel}}_2$ follows the Rician fading, and ${\rm{channel}}_3$ follows the Geometric fading.

    \textit{Geometric fading: All transceiver pairs are randomly distributed in an $R\times R$ area, as}
    \begin{equation}
        |h_{jk}|^2 = \frac{1}{1+d_{jk}^2}|r_{jk}|^2, \forall j,k,
    \end{equation}
    \textit{where $r_{jk}$ denotes the small-scale fading coefficient follows $\mathcal{CN}$(0, 1), $d_{jk}$ is the distance between the $j$-th transmitter and the $k$-th receiver.}

The data generation procedure is the same as that in Section \ref{beamforming_sim} and we again normalize the sum rate by using the WMMSE algorithm in order to compare the sum-rate performance under different channel distributions, which is expressed as the ‘Normalized Sumrate’ in the following simulation results. The system parameters and the network parameters are summarized in Table \ref{table6} and Table \ref{table7}, respectively.

\begin{table}[htp]
\vspace{-0.5cm}
    \renewcommand\arraystretch{1}
	\small
	\caption{System Parameters}
	\label{table6}
	\centering
	\scalebox{1}{
	\begin{tabular}{c|c}
		\toprule
		\textbf{Parameter} & \textbf{Value}\\
		\hline\hline
		Transceiver pairs, $K$ & $10$\\
		\hline
		Noise power, $\sigma^2$& $-10$ dB\\
		\hline
		Maximum transmit power, $P_{\rm{max}}$& $1$ w\\
		\hline
		Area length, $R$& $10$ m\\
		\bottomrule
	\end{tabular}}
	\vspace{-0.3cm}
\end{table}
\begin{table}[htp]
    \renewcommand\arraystretch{1}
	\small
	\caption{Neural Network Parameters}
	\label{table7}
	\centering
	\scalebox{0.92}{
	\begin{tabular}{c|c}
		\toprule
		\textbf{Parameter} & \textbf{Value}\\
		\hline\hline
		Type of Neural Network & CNN\\
		\hline
		\makecell{Number of layers in outer and\\ inner networks} & $2$, $2$\\
		\hline
		\makecell{Number of channels in outer and \\inner networks} & \makecell{\{$1, 6$\}\{$6, 8$\},\\\{$1, 4$\}\{$4, 8$\}}\\
		\hline
		Kernel size in outer and inner networks & $3 \times 3$, $3 \times 3$\\
		\hline
		Stride and padding in the convolution layer & $1$, $0$\\
		\hline
		Nonlinear function, $\sigma(x)$ & $\sigma(x)=\frac{1}{1+exp(-x)}$\\
		\bottomrule
	\end{tabular}}
\end{table}
\subsubsection{Three Important Goals} 
\begin{figure}[htp]
\centering
\vspace{-0.3cm}
\includegraphics[width=8cm]{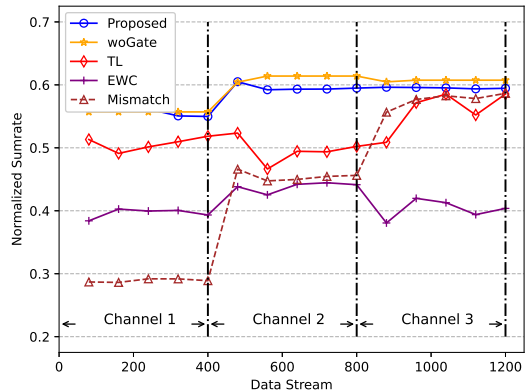}
\vspace{-0.5cm}
\caption{Sum-rate performances with different methods.}
\label{fig: sumrate_CNN}
\end{figure}

(\romannumeral1) \textbf{Seamlessness}. Fig. \ref{fig: sumrate_CNN} compares the proposed meta-gating framework with other benchmark algorithms on the sum-rate performance. The proposed meta-gating framework has the best sum-rate performance on each channel distribution compared to the benchmark algorithms due to its better model initializations. Table \ref{table8} presents the variance of the sum-rate performance among each channel distribution with different methods. It can be observed that the proposed framework has quite small variance, which indicates that the proposed framework can achieve similar sum-rate performances on different channel distributions. Therefore, it can well achieve the goals of ‘seamlessness’. These results are similar as those observed in Fig. \ref{fig: sumrate} and Table \ref{table3}.  
\begin{table}[htp]
\vspace{-0.6cm}
    \renewcommand\arraystretch{1}
	\small
	\caption{Variances of the normalized sum-rate performances under different methods.}
	\label{table8}
	\centering
	\scalebox{0.96}{
	\begin{tabular}{c| c| c}
		\toprule
		\textbf{Method} & (\textbf{Channel 1}, \textbf{Channel 2}) &(\textbf{Channel 2}, \textbf{Channel 3})\\
		\hline
		TL&$2.88\times 10^{-5}$&$1.04\times 10^{-3}$\\
		\hline
		EWC&$4.47\times 10^{-4}$&$3.26\times 10^{-4}$\\
		\hline
		Joint&$4.13\times 10^{-3}$&$1.34\times 10^{-3}$\\
		\hline
		WoGate&$7.57\times 10^{-4}$&$6.65\times 10^{-6}$\\
		\hline
		Proposed&$3.59\times 10^{-4}$&$9.76\times 10^{-8}$\\
		\bottomrule
	\end{tabular}}
\end{table}

\begin{figure}[htp]
\centering
\begin{minipage}[t]{0.48\textwidth}
\centering
\includegraphics[width=8cm]{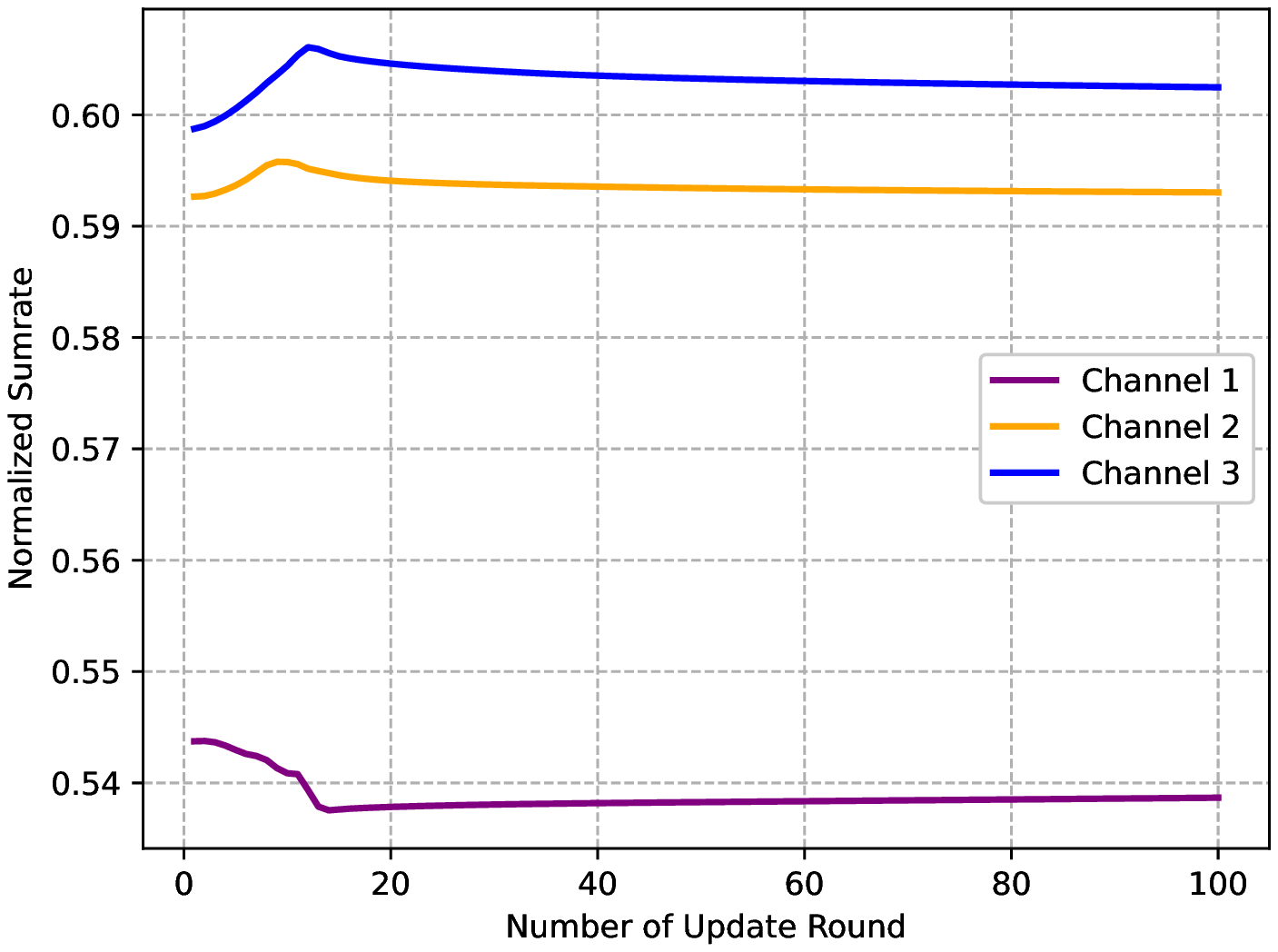}
\vspace{-0.5cm}
\caption{Sum-rate performances with different values of $J_q$.}
\label{fig: sumrate_diff_q_CNN}
\vspace{0.5cm}
\end{minipage}
\begin{minipage}[t]{0.48\textwidth}
\centering
\includegraphics[width=8cm, height=7cm]{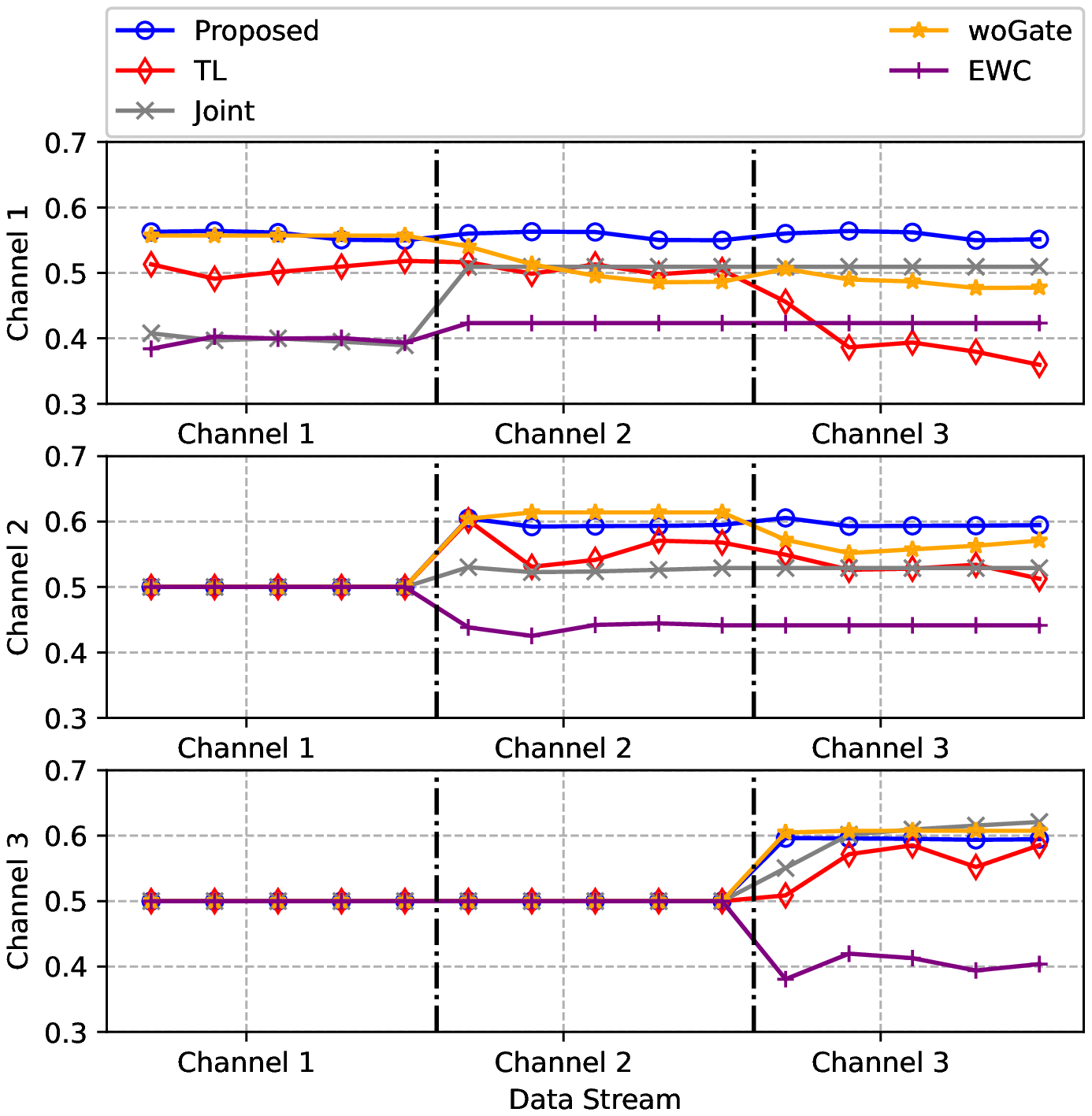}
\vspace{-0.5cm}
\caption{Capability for continuous adaptation of different methods.}
\label{fig: notforgetting_CNN}
\end{minipage}
\vspace{-0.2cm}
\end{figure}

(\romannumeral2) \textbf{Quickness}. Fig. \ref{fig: sumrate_diff_q_CNN} depicts the impact of different values of $J_q$ on the sum-rate performance, where the similar conclusion can be concluded as those in Fig. \ref{fig: sumrate_diff_q}.

(\romannumeral2) \textbf{Continuity}. Fig. \ref{fig: notforgetting_CNN} depicts the capability for continuous adaptation of different methods and we set the value of the normalized sum rate in ${\rm{channel}}_j$ as $0.5$ when $j \textgreater i$. Similarly, we compute the similarity among these three channels, which can be seen in Table \ref{table9}. From the table, the distance between ${\rm{channel}}_1$ and ${\rm{channel}}_3$ is the largest. Therefore, it would cause a large performance loss in ${\rm{channel}}_1$ when the model is updated according to the samples in ${\rm{channel}}_3$ in TL, as shown in Fig. \ref{fig: notforgetting_CNN}.

\begin{table}[htp]
 \renewcommand\arraystretch{1.3}
	\normalsize
	\vspace{-0.3cm}
	\caption{Distance between each channel}
	\label{table9}
	\centering
	\scalebox{0.9}{
	\begin{tabular}{c c c c}
		\toprule
		Channel pair &(1,2)&(2,3)&(3,1)\\
		\hline
		CDS&$0.046$&$-0.003$&$-0.101$\\
		\bottomrule
	\end{tabular}}
	\vspace{-0.2cm}
\end{table}

To conclude, simulation results demonstrate that the proposed framework is model-agnostic, i.e., it can well achieve three goals compared with the state-of-the-art algorithms on the proposed problem for both CNN and GNN models.

\subsubsection{Generalization Ability}
To verify Theorem \ref{theo 2} with simulation experiments, we test the sum-rate performances on the Nakagami-$m$ channels. The reason why we apply the Nakagami-$m$ channel is that it is a more general fading channel model and the Nakagami fading can be transformed into a variety of fading models by changing the value of $m$ (e.g., it can be degenerated into Rayleigh fading when $m=1$). Specifically, we randomly generate four kinds of channels where $|h_{j,k}|\sim Nakagami(m, \Omega)$ and $m$ follows a uniform distribution between $[0.5, 2]$ in the training stage. Similarly, we generate four kinds of channels in the testing stage, where the latter two kinds of channels are unseen channels following the Nakagami-$m$ distribution with different values of $m$ from the training stage. 

\begin{figure}[htp]
    \centering
    \vspace{-0.5cm}
    \includegraphics[width=7cm]{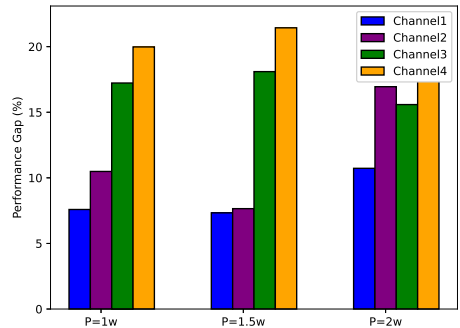}
    \vspace{-0.6cm}
    \caption{Sum-rate performance gaps between the proposed framework and the joint training method under different values of $P_{\rm{max}}$.}
    \label{fig: gengap}
\end{figure}

Table \ref{table11} presents the sum-rate performances of joint training method and proposed framework in both seen and unseen channels with different values of $P_{\rm{max}}$. It is observed that the proposed framework achieves a similar sum-rate performance with the joint training method on seen channels but significantly outperforms the joint training method on unseen channels, where the sum-rate performance gaps are further depicted in Fig. \ref{fig: gengap}. It is mainly because that the joint training method only focuses on the distribution on the seen channels, but the proposed framework can better take further optimizations into account because of the dual-loop optimization for better model initializations. 

\begin{table}[htp]
\vspace{-0.3cm}
	\small
    \renewcommand\arraystretch{1}
    \setlength\tabcolsep{8pt}
	\caption{Average normalized sum-rate performances of the proposed framework and the joint method under different values of $P_{\rm{max}}$.}
	\label{table11}
	\centering
	\scalebox{0.8}{
	\begin{tabular}{c|c|c|c|c|c}
		\toprule
		\multicolumn{2}{c|}{\multirow{2}*{$P_{\rm{max}}$/w}} & \multicolumn{2}{c|}{Seen} & \multicolumn{2}{c}{Unseen}\\
		\cline{3-6}
		\multicolumn{2}{c|}{~} & Channel 1 & Channel 2 & Channel 3 & Channel 4 \\
		\hline
		\rowcolor{gray!30}\cellcolor{white}{\multirow{2}*{$1$}} &Proposed& $0.6023$ & $0.6411$ & $0.6503$ & $0.6985$\\
		\cline{2-6}
		~ &Joint& $0.5580$ & $0.5802$ & $0.5547$ & $0.5822$\\
		\hline
		\rowcolor{gray!30}\cellcolor{white}\multirow{2}*{$1.5$} &Proposed& $0.5257$ & $0.5638$ & $0.5743$ & $0.6192$\\
		\cline{2-6}
		~ &Joint& $0.4898$ & $0.5237$ & $0.4863$ & $0.5099$\\
		\hline
		\rowcolor{gray!30}\cellcolor{white}\multirow{2}*{$2$} &Proposed& $0.4880$ & $0.5235$ & $0.5353$ & $0.5801$\\
		\cline{2-6}
		~ &Joint& $0.4408$ & $0.4476$ & $0.4631$ & $0.4755$\\
		\bottomrule
	\end{tabular}}
\end{table}

\section{Conclusions}\label{conclusion}
\indent In this paper, we have proposed a general meta-gating framework for solving wireless resource allocation problems in an episodically dynamic wireless environment, where the CSI distribution changes over periods and remains constant within each period. Specifically, the proposed framework includes an inner network and an outer network, and they are connected through the gating operation. The proposed dual-loop training method is developed to achieve the goals of ‘seamlessness’ and ‘quickness’ by combining the MAML algorithm with the unsupervised training method. As for the goal of ‘continuity’, the outer network learns to evaluate the importance of inner network’s parameters under different CSI distributions and then decide which subset of the inner network should be activated. Therefore, it enables the selective plasticity of the inner network. Additionally, we have theoretically analyzed the performance of the proposed meta-gating framework. Finally, simulation results have demonstrated that the proposed meta-gating framework can well adapt to the dynamic wireless environment via achieving three important goals compared with several existing state-of-the-art algorithms.\\
\newpage
\appendices
\section{Proof of Theorm 1} \label{appendix_a}

\begin{lemma}\label{lemma2}
\emph{Assume that function $\mathcal{L}(\cdot)$ is $L$-smooth in $\bm{\theta}_c$. If $\beta\leq\frac{1}{L}$, then it holds for any channel $c$ and any parameters $\bm{\theta}_c$ that}
\begin{align}
    &\mathcal{L}_{D_c}(\bm{\theta}_c^{J_q},\bm{\phi}^*)-\mathcal{L}_{D_c}(\bm{\theta}_c^1,\bm{\phi}^*)\leq \frac{1}{2\beta}\Vert\bm{\theta}_c-\bm{\theta}^*\Vert^2-\notag\\&\beta(1-\frac{\beta L}{2})\sum_{t=1}^{J_q-1}\Vert\triangledown\mathcal{L}_{D_c}(\bm{\theta}_c^t|\bm{\phi}^*)\Vert_2^2,
\end{align}
\emph{where $\bm{\theta}^*$ denotes the learned initializations of the inner network and $\bm{\theta}^{J_q}_c=\bm{\theta}^*-\beta\left(\triangledown\mathcal{L}_{D_c}(\bm{\theta}^*, \bm{\phi}^*)+\sum_{t=1}^{J_q-1}\triangledown\mathcal{L}_{D_c}(\bm{\theta}_c^t, \bm{\phi}^*)\right)$}.
\end{lemma}

\begin{lemma}\label{lemma4}
\emph{Assume that function $\mathcal{L}(\cdot)$ is $L$-smooth in $\bm{\theta}_c$. We denote the expected and empirical losses on $D_c$ as $\mathcal{L}(\bm{\theta}_c)$ and $\mathcal{L}_{D_c}(\bm{\theta}_c)$, respectively. Here $\beta\leq\frac{1}{L}$, given a channel $c$, considering the empirical minimization problem in (\ref{mini_problem})-(\ref{mini_problem_1}).}
\end{lemma}
\begin{figure*}
\begin{align}
    \bm{\theta}_c^1 &= \mathop{\rm{argmin}}\limits_{\bm{\theta}_c}\left\{h_{D_c}(\bm{\theta}_c)=\left<\triangledown\mathcal{L}_{D_c}(\bm{\theta}^*|\bm{\phi}^*) +\sum_{t=1}^{J_q-1}\Vert\triangledown\mathcal{L}_{D_c}(\bm{\theta}_c^t|\bm{\phi}^*), \bm{\theta}_c-\bm{\theta}^*\right>+\frac{1}{2\beta}\Vert\bm{\theta}_c-\bm{\theta}^*\Vert_2^2\right\}, \label{mini_problem}\\ 
    &= \bm{\theta}^*-\beta\left(\triangledown\mathcal{L}_{D_c}(\bm{\theta}^*|\bm{\phi}^*) + \sum_{t=1}^{J_q-1}\triangledown\mathcal{L}_{D_c}(\bm{\theta}_c^t|\bm{\phi}^*)\right).\label{mini_problem_1}
\end{align}
{\noindent} \rule[-10pt]{18cm}{0.05em}
\end{figure*}
\emph{Then, we can obtain the following bound for any $J_q$}
\begin{equation}
    \vert\mathbb{E}_{D_c\sim \mathcal{C}}[\mathcal{L}(\bm{\theta}_c^{J_q})-\mathcal{L}_{D_c}(\bm{\theta}_c^{J_q})]\vert\leq\frac{2G^2[(1+2\beta L)^{J_q}-1]}{LN_m^{te}}.
\end{equation}

The proof of the aforementioned two lemmas can be found in \cite{supplementary}.

Then, we can obtain the following upper bound according to Lemma \ref{lemma4} as follows:

\begin{align}
    &\mathbb{E}_{D_c}\left[\mathcal{L}(\bm{\theta}_c^{J_q}, \bm{\phi}^*)-\mathcal{L}(\bm{\theta}^*, \bm{\phi}^*)\right] \notag\\
    &=\mathbb{E}_{D_c}\left[\mathcal{L}(\bm{\theta}_c^{J_q}, \bm{\phi}^*)-\mathcal{L}_{D_c}(\bm{\theta}^{J_q}_c, \bm{\phi}^*)\right] \notag\\
    &+ \mathbb{E}_{D_c}\left[\mathcal{L}_{D_c}(\bm{\theta}^{J_q}_c, \bm{\phi}^*)-\mathcal{L}(\bm{\theta}_c^*, \bm{\phi}^*)\right],\\
    &\leq\vert\mathbb{E}_{D_c}\left[\mathcal{L}(\bm{\theta}_c^{J_q}, \bm{\phi}^*)-\mathcal{L}_{D_c}(\bm{\theta}^{J_q}_c, \bm{\phi}^*)\right]\vert\notag\\
    &+ \mathbb{E}_{D_c}\left[\mathcal{L}_{D_c}(\bm{\theta}^{J_q}_c, \bm{\phi}^*)-\mathcal{L}(\bm{\theta}_c^*, \bm{\phi}^*)\right],\\
    &\leq\frac{2G^2[(1+2\beta L)^{J_q}-1]}{LN_m^{te}} + \mathbb{E}_{D_c}\left[\mathcal{L}_{D_c}(\bm{\theta}^{J_q}_c, \bm{\phi}^*)-\mathcal{L}(\bm{\theta}_c^*, \bm{\phi}^*)\right].\label{final}
\end{align}

Note that $\mathbb{E}_{D_c}\left[\mathcal{L}_{D_c}(\bm{\theta}_c^*, \bm{\phi}^*)\right] = \mathbb{E}_{D_c}\left[\mathcal{L}(\bm{\theta}_c^*, \bm{\phi}^*)\right]$. Then, we take the expectation on both sides of (\ref{final}) on $c\sim\mathcal{C}$ to finally obtain
\begin{align}
    &\mathbb{E}_{c\sim\mathcal{C}}\mathbb{E}_{D_c}\left[\mathcal{L}(\bm{\theta}_c^{J_q}, \bm{\phi}^*)-\mathcal{L}(\bm{\theta}^*, \bm{\phi}^*)\right]
    \leq\frac{2G^2[(1+2\beta L)^{J_q}-1]}{LN_m^{te}}\\
    &+ \mathbb{E}_{D_c}\left[\mathcal{L}_{D_c}(\bm{\theta}^{J_q}_c, \bm{\phi}^*)-\mathcal{L}(\bm{\theta}_c^*, \bm{\phi}^*)\right].\notag
\end{align}

\section{Proof of Theorm 2} \label{appendix_b}

Consider a fixed channel $c\sim\mathcal{C}$ and its associated random dataset $D_c\sim c$ with size $N_m^{te}$. Then, we perform $J_q$ gradient steps to obtain the adapted parameter $\bm{\theta}^{J_q}_c=\bm{\theta}^*-\beta\left(\triangledown\mathcal{L}_{D_c}(\bm{\theta}^*|\bm{\phi}^*)+\sum_{t=1}^{J_q-1}\triangledown\mathcal{L}_{D_c}(\bm{\theta}_c^t|\bm{\phi}^*)\right)$. We can show the following inequality:
\begin{align}
    \Vert\mathbb{E}_{D_c}&[\triangledown\mathcal{L}(\bm{\theta}_c^{J_q}|\bm{\phi}^*)]\Vert^2 = \Vert\mathbb{E}_{D_c}[\triangledown\mathcal{L}(\bm{\theta}_c^{J_q}|\bm{\phi}^*)-\triangledown\mathcal{L}_{D_c}(\bm{\theta}_c^{J_q}|\bm{\phi}^*)] \notag\\
    &+ \mathbb{E}_{D_c}[\triangledown\mathcal{L}_{D_c}(\bm{\theta}_c^{J_q}|\bm{\phi}^*)]\Vert^2,\\
    &\leq 2\Vert\mathbb{E}_{D_c}[\triangledown\mathcal{L}(\bm{\theta}_c^{J_q}|\bm{\phi}^*)-\triangledown\mathcal{L}_{D_c}(\bm{\theta}_c^{J_q}|\bm{\phi}^*)]\Vert^2 \notag\\
    &+ 2\Vert\mathbb{E}_{D_c}[\triangledown\mathcal{L}_{D_c}(\bm{\theta}_c^{J_q}|\bm{\phi}^*)]\Vert^2,\\
    &\leq 2\Vert\mathbb{E}_{D_c}[\triangledown\mathcal{L}(\bm{\theta}_c^{J_q}|\bm{\phi}^*)-\triangledown\mathcal{L}_{D_c}(\bm{\theta}_c^{J_q}|\bm{\phi}^*)]\Vert^2 \notag\\ &+2\mathbb{E}_{D_c}\left[\Vert\triangledown\mathcal{L}_{D_c}(\bm{\theta}_c^{J_q}|\bm{\phi}^*)\Vert^2\right],\\
    &\mathop{\leq}^{\tiny\textcircled{1}} \frac{8G^2[(1+2\beta L)^{J_q}-1]^2}{{N_m^{te}}^2} +2\mathbb{E}_{D_c}\left[\Vert\triangledown\mathcal{L}_{D_c}(\bm{\theta}_c^{J_q}|\bm{\phi}^*)\Vert^2\right].
\end{align}
Here, {\tiny\textcircled{1}} comes from Lemma \ref{lemma4}.

\end{document}